\documentclass[
	reprint,
	article,
	twocolumn,
	nofootinbib,
	nobibnotes,
	floatfix,
	superscriptaddress
]{revtex4-2}

\usepackage{amssymb}
\usepackage{amsmath}
\usepackage{graphicx}
\usepackage{dcolumn}
\usepackage{bm}
\usepackage{hyperref}
\usepackage[usenames,dvipsnames]{xcolor}
\usepackage{listings}
\usepackage[utf8]{inputenc}
\usepackage[english]{babel}
\usepackage{booktabs}
\usepackage{hhline}

\definecolor{LightGray}{gray}{0.95}
\definecolor{pycommentcol}{rgb}{0.35,0.35,0.35}  % gray
\definecolor{pystatecol}{rgb}{0,0,0.7}           % blue
\definecolor{pystringcol}{rgb}{0,0.6,0}          % green
\definecolor{pyinbuiltscol}{rgb}{0.55,0.15,0.55} % plum
\definecolor{pyspecialcol}{rgb}{0.8,0.45,0.12}   % orange
\definecolor{pynumbercol}{rgb}{0.6, 0.0, 0.0}    % red

\newcommand*{\pyfontfamily}{\fontfamily{DejaVuSansMono-TLF}\selectfont}

\usepackage{csquotes}
 
\newcommand{\amep}{\texttt{AMEP}}

\newcommand\diff{\mathrm{d}}

\newcommand\pythonstyle{\lstset{
    language=Python,
    basicstyle=\pyfontfamily,
    commentstyle=\color{pycommentcol}\itshape,
    emph={self,cls,@classmethod,@property}, % Custom highlighting
    emphstyle=\color{pyspecialcol}\itshape, % Custom highlighting style
    morestring=[b]{"""},
    stringstyle=\color{pystringcol}\itshape,
    keywordstyle=\color{pystatecol}\bfseries,        % statements
    deletekeywords={print,reload},
    deletekeywords=[2]{sum,reload},
    classoffset=1,
    morekeywords={print,None,TypeError,True,False},
    keywordstyle=\color{pyinbuiltscol},
    frame=tb,                        
    showstringspaces=false,
    backgroundcolor=\color{LightGray},
    numbers=left,                    
    numbersep=5pt,
    numberstyle=\tiny\color{pycommentcol},
    breaklines=true,
    literate=*
    {0}{{{\color{pynumbercol}0}}}{1}
    {1}{{{\color{pynumbercol}1}}}{1}
    {2}{{{\color{pynumbercol}2}}}{1}
    {3}{{{\color{pynumbercol}3}}}{1}
    {4}{{{\color{pynumbercol}4}}}{1}
    {5}{{{\color{pynumbercol}5}}}{1}
    {6}{{{\color{pynumbercol}6}}}{1}
    {7}{{{\color{pynumbercol}7}}}{1}
    {8}{{{\color{pynumbercol}8}}}{1}
    {9}{{{\color{pynumbercol}9}}}{1}
}}

\lstnewenvironment{python}[1][]
{
    \pythonstyle
    \scriptsize
    \lstset{#1}
}
{}
\newcommand\pythoninline[1]{{\scriptsize\pythonstyle\lstinline!#1!}}

\newcommand\pythonoutstyle{\lstset{
    basicstyle=\pyfontfamily,
    showstringspaces=false
}}
\lstnewenvironment{pythonout}[1][]
{
    \pythonoutstyle
    \scriptsize
    \lstset{#1}
}
{}
\newcommand\pythonoutinline[1]{{\pythonoutstyle\lstinline!#1!}}

\lstdefinestyle{command}{
    backgroundcolor=\color{black},   
    commentstyle=\color{white},
    keywordstyle=\color{white},
    numberstyle=\tiny\color{pycommentcol},
    stringstyle=\color{white},
    basicstyle=\ttfamily\footnotesize\color{white},
    breakatwhitespace=false,         
    breaklines=true,                 
    captionpos=b,                    
    keepspaces=true,                 
    numbers=left,                    
    numbersep=5pt,                  
    showspaces=false,                
    showstringspaces=false,
    showtabs=false,                  
    tabsize=2
}

\begin{document}

\title[AMEP: The Active Matter Evaluation Package for Python]{AMEP: The Active Matter Evaluation Package for Python}

\author{Lukas Hecht}
\email{lukas.hecht@pkm.tu-darmstadt.de}
\affiliation{ 
	Institute for Condensed Matter Physics, Department of Physics, Technische Universit\"at Darmstadt, Hochschulstr.\ 8, 64289 Darmstadt, Germany
}
\author{Kay-Robert Dormann}
\affiliation{ 
	Institute for Condensed Matter Physics, Department of Physics, Technische Universit\"at Darmstadt, Hochschulstr.\ 8, 64289 Darmstadt, Germany
}
\author{Kai Luca Spanheimer}
\affiliation{ 
	Institute for Condensed Matter Physics, Department of Physics, Technische Universit\"at Darmstadt, Hochschulstr.\ 8, 64289 Darmstadt, Germany
}
\author{Mahdieh Ebrahimi}
\affiliation{ 
	Institute for Condensed Matter Physics, Department of Physics, Technische Universit\"at Darmstadt, Hochschulstr.\ 8, 64289 Darmstadt, Germany
}
\author{Malte Cordts}
\affiliation{ 
	Institute for Condensed Matter Physics, Department of Physics, Technische Universit\"at Darmstadt, Hochschulstr.\ 8, 64289 Darmstadt, Germany
}
\author{Suvendu Mandal}
\affiliation{ 
	Institute for Condensed Matter Physics, Department of Physics, Technische Universit\"at Darmstadt, Hochschulstr.\ 8, 64289 Darmstadt, Germany
}
\author{Aritra K.\ Mukhopadhyay}
\affiliation{ 
	Institute for Condensed Matter Physics, Department of Physics, Technische Universit\"at Darmstadt, Hochschulstr.\ 8, 64289 Darmstadt, Germany
}
\author{Benno Liebchen}
\email{benno.liebchen@pkm.tu-darmstadt.de}
\affiliation{ 
	Institute for Condensed Matter Physics, Department of Physics, Technische Universit\"at Darmstadt, Hochschulstr.\ 8, 64289 Darmstadt, Germany
}

\date{\today}

% =================================================================================================================
% ABSTRACT
% =================================================================================================================
\begin{abstract}
The Active Matter Evaluation Package (\amep) is a Python library for analyzing simulation data of particle-based and continuum simulations. It provides a powerful and simple interface for handling large data sets and for calculating and visualizing a broad variety of observables that are relevant to active matter systems. Examples range from the mean-square displacement and the structure factor to cluster-size distributions, binder cumulants, and growth exponents. \amep\ is written in pure Python and is based on powerful libraries such as NumPy, SciPy, Matplotlib, and scikit-image. Computationally expensive methods are parallelized and optimized to run efficiently on workstations, laptops, and high-performance computing architectures, and an HDF5-based data format is used in the backend to store and handle simulation data as well as analysis results. \amep\ provides the first comprehensive framework for analyzing simulation results of both particle-based and continuum simulations (as well as experimental data) of active matter systems. In particular, \amep\ also allows it to analyze simulations that combine particle-based and continuum techniques such as used to study the motion of bacteria in chemical fields or for modeling particle motion in a flow field. \amep\ is available at \url{https://amepproject.de} and can be installed via conda and pip.
\end{abstract}

\maketitle

% =================================================================================================================
% =================================================================================================================
% MAIN TEXT
% =================================================================================================================
% =================================================================================================================

% =================================================================================================================
% INTRODUCTION
% =================================================================================================================
\section{Introduction}
\label{sec:Introduction}
Computer simulations are a powerful method to investigate and understand the physical properties of soft matter and biological systems. In particular, molecular dynamics (MD) simulations stand out as an indispensable tool to determine the microscopic dynamics and structural properties of molecular systems comprising bio-molecules \cite{Karplus_NatStructBiol_2002,Kumarietal_CurrProtPeptSci_2017,Marwick_PhysChemChemPhys_2011}, polymer electrolytes \cite{Mogurampelly_AnnRevChemBioEng_2016,Yao_ChemRev_2022,Haskins_JPhysChem_2014}, liquid crystals \cite{Wilson_IntRevPhysChem_2007,Huang_SoftMatter_2023,Watanabe_Symmetry_2023}, or confined liquids \cite{Horstmann_Langmuir_2022,Pal_JPhysChemC_2018,Mandal_NatComm_2014,MandalSuvendu_PhysRevLett_2017} for example. These systems exhibit notable relevance in both industrial and medical applications \cite{Guarra_JChemTheoryComput_2023,Rafi_JBioStructDyn_2020,Smith_Processes_2022}. Expanding beyond atomistic and systematically coarse-grained MD simulations involving suitable force fields, Brownian dynamics (BD) simulations have extensively been used over the past two decades especially also for modelling active matter systems. In such systems, the individual constituents consume energy from their environment and use it to self propel and navigate through complex surroundings \cite{Magistris_PhysicaA_2015,Bechinger_RevModPhys_2016}. Examples of active matter systems can be observed across scales from microscopic entities such as bacteria, algae, and synthetic microswimmers \cite{Elgeti_RepProgPhys_2015,Liu_PhysRevLett_2019,Wolgemuth_CurrBiol_2002,Rafai_PhysRevLett_2010,Howse_PhysRevLett_2007,Liebchen_AccChemRes_2018,Qiao_PhysRevRes_2020,Kurzthaler_PhysRevLett_2024,Ramos_SoftMatter_2020,Kurzthaler_PhysRevLett_2018} to fishes, birds, drones, and human crowds on the macroscale \cite{Grauer_SciRep_2020,Attanasi_NatPhys_2014,Klotsa_SoftMatter_2019,Buhl_Science_2006,Hemelrijk_InterFoc_2012,Ballerini_PNAS_2008,Bialek_PNAS_2012,Silverberg_PhysRevLett_2013,Bain_Science_2019}. These out-of-equilibrium systems exhibit striking collective phenomena such as phase separation (where the system selects a density) \cite{Cates_AnnuRevCondensMatterPhys_2015,Barre_JStatPhys_2015,Gonnella_CRPhysique_2015,Buttinoni_PhysRevLett_2013,Blaschke_SoftMatter_2016,Digregorio_PhysRevLett_2018,Bergmann_PhysRevE_2018,Dolai_SoftMatter_2018,Zhang_Nature_2021,Su_NewJPhys_2021}, non-equilibrium pattern formation (where the system selects a length scale) \cite{Farrell_PhysRevLett_2012,Liebchen_PhysRevLett_2015,Liebchen_PhysRevLett_2017_2,Saintillan_PhysRevLett_2008,Solon_PhysRevE_2015,Hagan_CurrOpCellBiol_2016,Palacci_PhysRevLett_2010_2}, or other ordering transitions such as flocking showing (long-range) orientational ordering \cite{Toner_PhysRevE_1998,Bialek_PNAS_2012,Levis_PhysRevRes_2019,Chatterjee_PhysRevX_2021}.

To effectively model active matter systems, different computational approaches provide distinct advantages \cite{Shaebani_NatRevPhys_2020,Hecht_ArXiv_2021}. Particle-based models such as the active Brownian particle (ABP) model \cite{Romanczuk_EurPhysJSpecialTopics_2012}, solved numerically using BD simulations for example, have proven valuable for investigating collective phenomena such as motility-induced phase separation (MIPS) \cite{Digregorio_PhysRevLett_2018,DeKarmakar_SoftMatter_2022,Redner_PhysRevLett_2013,Cates_AnnuRevCondensMatterPhys_2015,Su_CommPhys_2023,Redner_PhysRevLett_2016,Takatori_PhysRevE_2015,Speck_PhysRevLett_2014,Tailleur_PhysRevLett_2008,Rojas-Vega_EurPhysJE_2023,Ma_JCP_2022,Nie_PhysRevRes_2020,Yang_PhysRevE_2022,Mandal_PhysRevLett_2019,Hecht_PRL_2022} as well as the dynamics and local order of the involved individual particles \cite{Digregorio_SoftMatter_2021,Elgeti_RepProgPhys_2015,Caporusso_PhysRevLett_2023,Saw_PhysRevE_2023,Omar_JCP_2023,Sprenger_JPhysCondensMatter_2023,Schiltz_Rouse_ArXiv_2023,Sandoval_SoftMatter_2023,Chen_EPL_2023}. Conversely, continuum models such as the active model B+ \cite{Tjhung_PhysRevX_2018} enable us to understand collective phenomena over larger length and time scales by studying the evolution of particle densities  \cite{Liebchen_PhysRevLett_2015,Tiribocchi_PhysRevLett_2015,Stenhammar_PhysRevLett_2013,Speck_PhysRevLett_2014,Tjhung_PhysRevX_2018,Mishra_PhysRevE_2010,Solon_PhysRevE_2015,Chatterjee_PhysRevX_2021,Zhao_PRL_2023,Berx_EPL_2023}. Therefore, it is a common approach to start from a particle-based model and subsequently derive a continuum model via coarse-graining techniques \cite{Levis_PhysRevRes_2019,Liebchen_PhysRevLett_2017_2,Sese-Sansa_SoftMatter_2022,Stenhammar_SoftMatter_2014,Menzel_JCP_2016,Hoell_JCP_2018,Hoell_NewJPhys_2017,Soto_ArXiv_2024,Dean_JPhysA_1996,Tailleur_PhysRevLett_2008,Risken_Book_TheFokker-PlanckEquation_1984}. Moreover, the integration of particle-based and continuum models has emerged as a promising approach \cite{Farrell_PhysRevLett_2012,Fily_PhysRevLett_2012,Grauer_SciRep_2020_2,Jiang_SoftMatter_2023,Liebchen_PhysRevLett_2017,Zampetaki_PNAS_2021,Liebchen_AccChemRes_2018,Fadda_SoftMatter_2023}, especially for scenarios such as modeling the interaction of active particles with a surrounding fluid or another medium that is quasi-continuous on the scale of the considered particles, as exemplified by the motion of bacteria or synthetic Janus colloids in a self-produced concentration field \cite{Fischer_JCP_2019,Grauer_SciRep_2020_2,Jiang_SoftMatter_2023,Liebchen_PhysRevLett_2017,Zampetaki_PNAS_2021,Saha_PhysRevE_2014,Pohl_PhysRevLett_2014}. These approaches are also particularly important for modeling artificial microswimmers such as active colloids in the presence of chemical fields \cite{Liebchen_JPhysCondensMatter_2022,Zoettl_AnnRevCondMatPhys_2023}, exhibiting phenomena such as chemotaxis \cite{Liebchen_ArXiv_2019,Keller_JTheorBiol_1971,Zhao_PRL_2023,Jiang_SoftMatter_2023}.

To gain physical insights from the data resulting from numerical solutions, a comprehensive array of analysis techniques is required. The mean-square displacement and time correlation functions such as the orientational autocorrelation function are popular observables to achieve insights into the dynamical aspects \cite{Sprenger_PhysRevE_2022,Nguyen_JPhysCondensMatter_2022,Lisin_PhysChemChemPhys_2022,Caprini_JCP_2021,Reichert_SoftMatter_2021,Sandoval_PhysRevE_2020,Breoni_PhysRevE_2020,Sprenger_JPhysCondensMatter_2023,Lemaitre_NewJPhys_2023,Debnath_JChemSci_2023,Nie_PhysRevE_2020}, while spatial correlation functions such as the radial/pair distribution function and the structure factor are frequently considered to provide information about spatial structures \cite{WangHaina_JCP_2020,Gavagnin_PhysRevE_2018,Dolai_SoftMatter_2018,Szamel_PhysRevE_2015,Broeker_SoftMatter_2024}. Other observables such as entropy production and kinetic temperature are also considered frequently to analyze the collective behavior of the system under investigation \cite{Frydel_PRE_2023,Mandal_PhysRevLett_2019,Hecht_PRL_2022,OByrne_NatRevPhys_2022,Ro_PhysRevLett_2022,GrandPre_PhysRevE_2021,Fodor_EPL_2021,Mandal_PhysRevLett_2017,Schiltz_Rouse_ArXiv_2023,DeKarmakar_PhysRevE_2020,Marconi_SciRep_2017,Cugliandolo_FluctNoiseLett_2019,Zhang_PhysRevRes_2023,Dabelow_JStatMech_2021,Cugliandolo_JPhysAMathTheor_2011,Szamel_PhysRevE_2019,Herpich_PhysRevE_2020,Gaspard_Research_2020,Venkatasubramanian_ComputChemEng_2022,Khali_PhysRevE_2024}, and together with statistical analysis tools such as binder cumulants used in finite size scaling analyses \cite{Digregorio_PhysRevLett_2018,Dittrich_EPJE_2021,Siebert_PhysRevE_2018,Gnan_SoftMatter_2022,Sinha_SoftMatter_2024} and cluster analyses including cluster growth exponents \cite{Mandal_PhysRevLett_2019,Wittkowski_NatComm_2014,Stenhammar_SoftMatter_2014,Caporusso_PhysRevLett_2023,Shi_PhysRevLett_2020,Su_NewJPhys_2021,Caporusso_PhysRevLett_2020,VanDerLinden_PhysRevLett_2019}, these observables allow us to characterize the non-equilibrium phase diagram and critical dynamics of active matter systems. Additionally, local order analyses such as local density and bond orientational order parameters offer insights into the system's local structure and symmetries \cite{Paoluzzi_CommPhys_2022,Redner_PhysRevLett_2013,Caprini_PhysRevRes_2020,Digregorio_SoftMatter_2021,DeKarmakar_SoftMatter_2022,Digregorio_PhysRevLett_2018,Klamser_JCP_2019,Rodriguez_SoftMatter_2020}. While existing analysis packages such as freud \cite{Ramasubramani_ComPhysComm_2020}, MDAnalysis \cite{Gowers_ProcPythonConf_2016,Michaud-Agrawal_JComputChem_2011}, VMD \cite{Humphrey_JMolGraph_1996}, MDTraj \cite{McGibbon_BiophysJ_2015}, Ovito \cite{Stukowski_ModellingSimulMaterSciEng_2009}, Pytim \cite{Sega_JComputChem_2018}, LOOS \cite{Romo_IEEE_2009,Romo_JComputChem_2014}, and MMTK \cite{Hinsen_JComputChem_2000} offer the possibility to calculate such observables, they are mostly inspired by and optimized for the MD simulation community and fall short in capturing all relevant observables for active matter systems in a single package. Additionally, they lack the possibility to handle continuum simulation data. Therefore, a new library is required that (i) consolidates the essential observables needed to analyze active matter simulation data, (ii) provides a unified platform for analyzing both particle-based and continuum simulation data, and (iii) seamlessly integrates data formats widely-used in computational physics through an application programming interface (API).

In this paper, we introduce the Active Matter Evaluation Package (\amep), a unified framework for analyzing MD simulation, BD simulation, and continuum simulation data with a specific focus on soft and active matter systems. \amep\ provides an optimized framework for loading, storing, and evaluating simulation data based on particle trajectories and the time evolution of continuum fields. This framework is based on an optimized HDF5 file format \cite{HDF5,Collette_Book_PythonAndHDF5_2013} which is optimal for long-term storage purposes and for handling data of large-scale simulations \cite{Collette_Book_PythonAndHDF5_2013,DeBuyl_CompPhysCommun_2014,Elhaddad_IWCCE_2015}. Computationally expensive methods are parallelized and \amep\ selectively loads only the data into the main memory which is necessary for the current computational step. This ensures efficient operation on high-performance computing architectures, workstations, and laptops. \amep\ is written in pure Python and provides a user-friendly, easy-to-learn Python API that interfaces with common tools used in computational physics via NumPy arrays \cite{Harris_Nature_2020,Walt_CompSciEng_2011}. Based on common Python libraries such as NumPy \cite{Harris_Nature_2020,Walt_CompSciEng_2011}, SciPy \cite{Virtanen_NatMeth_2020}, Matplotlib \cite{Hunter_SciProg_2007}, and scikit-image \cite{Walt_PeerJ_2014}, \amep\ provides a powerful toolbox for calculating spatial and temporal correlation functions, visualizing and animating simulation results, and coarse-graining particle-based simulations, which makes it possible to easily analyze the dynamics and structure of both particle-based and continuum simulation data.

The paper is organized as follows: We first demonstrate a minimal example on how to use \amep\ to load, analyze, and visualize simulation data in Section \ref{sec:Usage}. Second, we give a brief overview on the structure and design of \amep\ in Section \ref{sec:Design}. Finally, in Sections \ref{sec:Particles} and \ref{sec:Fields}, we apply a selection of analysis functions provided by \amep\ to particle-based simulations of the ABP model and to continuum simulations of the active model B+, respectively, and briefly discuss the results.

% =================================================================================================================
% HOW TO USE AMEP - MINIMAL EXAMPLE
% =================================================================================================================
\section{How to use \amep}
\label{sec:Usage}
\amep\ is designed with a user-focused mindset to simplify the access to simulation data within a few lines of Python code. Before discussing the general design of \amep\ and various examples on how to apply it to particle-based and continuum simulation data, we will demonstrate its general usage with a minimal example. For a quick start with \amep, we recommend to download the demo files at \url{https://github.com/amepproject/amep/tree/main/examples} and to run them part by part while reading this section.

\subsection{Exemplary data: The active Brownian particle model}
\label{sub:ABP-model}
Within the following example, we load simulation data from a particle-based simulation of the overdamped active Brownian particle (ABP) model in two spatial dimensions. The ABP model is one of the simplest and most popular models to describe active particles that self propel in a certain direction which smoothly changes due to rotational diffusion \cite{Romanczuk_EurPhysJSpecialTopics_2012,Hecht_ArXiv_2021,Mandal_PhysRevLett_2019,Loewen_JCP_2020,Su_NewJPhys_2021,Sandoval_PhysRevE_2020,Digregorio_SoftMatter_2021,Caporusso_PhysRevLett_2023,Redner_PhysRevLett_2013,Omar_PhysRevLett_2021}. Within this model, each particle is represented by a (slightly soft) disk/sphere of diameter $\sigma$, mass $m$, and moment of inertia $I$ and possesses an inherent internal drive, denoted by an effective self-propulsion force $\vec{F}_{\text{SP},i}=\gamma_{\rm t} v_0 \hat{p}(\theta_i)$, where the self-propulsion direction is represented by $\hat{p}(\theta_i)=(\cos \theta_i, \sin \theta_i)$ and $v_0$ denotes the terminal self-propulsion speed. The interactions between the particles are governed by an excluded-volume repulsive interaction potential $u$. The evolution of the particles' positions $\vec{r}_i$ and orientations $\theta_i$ adheres to the Langevin equations \cite{Loewen_JCP_2020,Sandoval_PhysRevE_2020,Mandal_PhysRevLett_2019,Hecht_PRL_2022,Khali_PhysRevE_2024}
\begin{align}   
	m \frac{\diff^2 \vec{r}_i}{\diff t^2} &= -\gamma_{\rm t} \frac{\diff \vec{r}_i}{\diff t} - \sum_{\substack{j=1\\j\neq i}}^{N}\nabla_{\vec{r}_i}u\left(r_{ji}\right) + \gamma_{\rm t} v_0 \hat{p}(\theta_i)\nonumber\\
	& + \sqrt{2 k_{\rm B} T_{\rm b} \gamma_{\rm t}} \vec{\eta}_i (t), \label{eq:abps-trans}\\ 
	I \frac{\diff^2 \theta_i}{\diff t^2} &= -\gamma_{\rm r} \frac{\diff \theta_i}{\diff t} + \sqrt{2k_{\rm B} T_{\rm b} \gamma_{\rm r}} \xi_i(t). \label{eq:abps-rot}
\end{align}
Here, $\vec{\eta}_i$ and $\xi_i$ denote Gaussian white noise with zero-mean and unit variance, $T_{\rm b}$ represents the bath temperature, $\gamma_{\rm t}$ and $\gamma_{\rm r}$ denote the translational and rotational drag coefficients, respectively, $k_{\rm B}$ is the Boltzmann constant, and $N$ denotes the total number of active particles. The overdamped limit can be obtained using $m/\gamma_{\rm t}\rightarrow 0$ and $I/\gamma_{\rm r}\rightarrow 0$ (see also Section \ref{sec:Particles}).

\subsection{Installing \amep}
\label{sub:installation}
\amep\ is a Python library that requires Python 3.10 or higher. To use \amep, we recommend to install Python via Anaconda\footnote{See \url{https://docs.anaconda.com/free/anaconda/install/index.html} for instructions on how to install Anaconda.}. If Anaconda is installed, one can create a new environment, activate it, and install \amep\ through the terminal (Linux/macOS) or the Anaconda Prompt (Windows):
\begin{lstlisting}[language=bash, style=command] 
conda create -n amepenv python=3.10
conda activate amepenv
conda install conda-forge::amep
\end{lstlisting}
Now, one can start the Python interpreter to import and use \amep:
\begin{lstlisting}[language=bash, style=command, firstnumber=4] 
python
>>> import amep
>>> 
\end{lstlisting}
Alternatively, we recommend to use Jupyter notebooks\footnote{For more information about Jupyter notebooks, see \url{https://jupyter.org/}.}.

\subsection{Analyzing simulation data using \amep}
\label{sub:how-to-use-example}
Now, we use \amep\ to calculate the mean-square displacement (MSD), the diffusion coefficient, and the orientation autocorrelation function (OACF) of a single ABP and the radial distribution function (RDF) of multiple interacting ABPs. At the end of this section, we visualize the results in a combined plot. Exemplary code and data of the relevant ABP simulations is available at \url{https://github.com/amepproject/amep/tree/main/examples}.

\begin{figure*}
	\centering
	\includegraphics[width=1.0\linewidth]{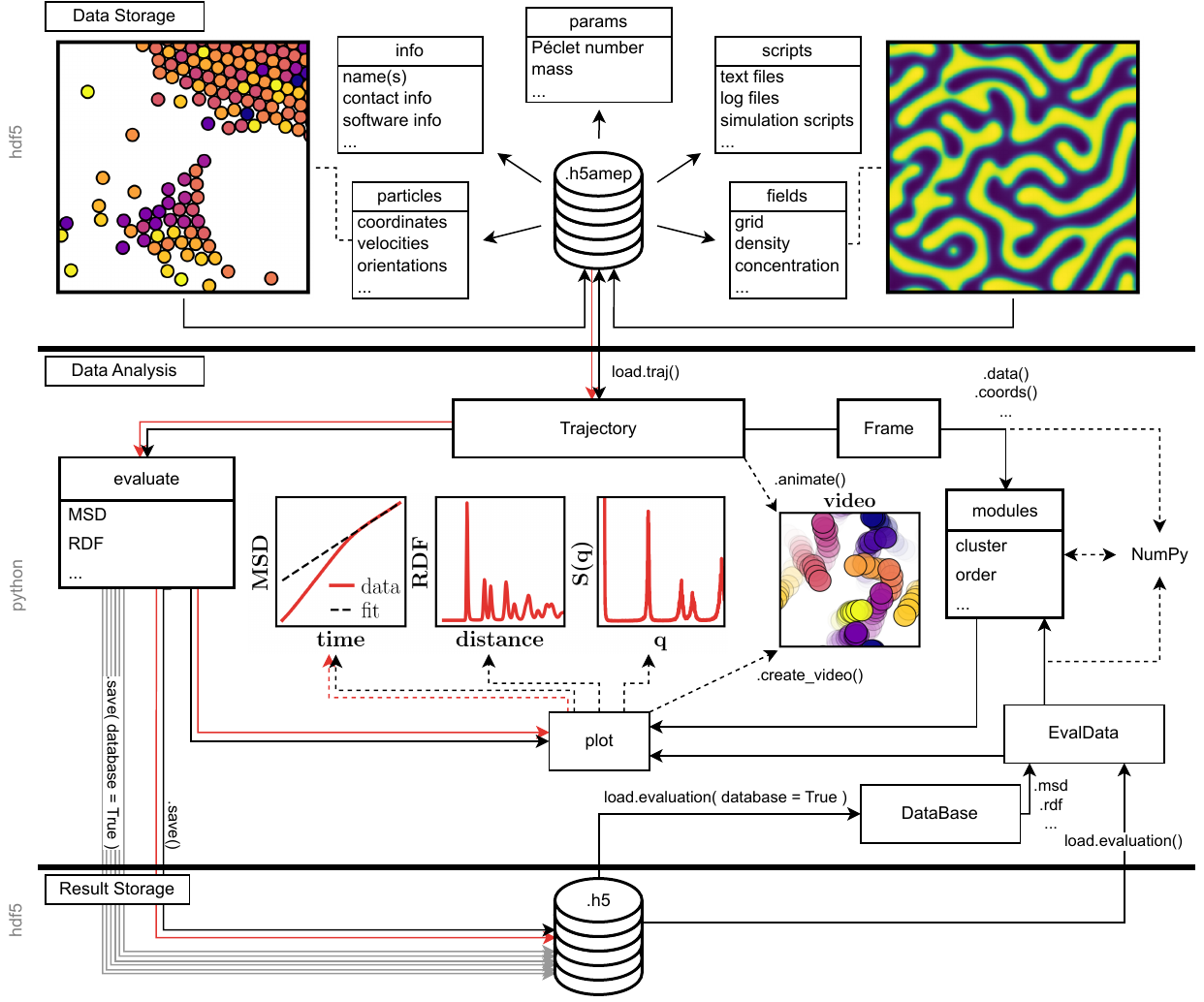}
	\caption{\textbf{\amep\ flow chart: Schematic of the design of \amep.} The first layer ``Data Storage'' represents the loading and storage of data from particle-based or continuum simulations, which is stored in HDF5 files (\texttt{.h5amep}) together with various metadata the user can add. The central part of the Python interface (layer ``Data Analysis'') is the \texttt{Trajectory} object which acts as a list of \texttt{Frame} objects and gives access to the data stored in the HDF5 file. The \texttt{Trajectory} object can be used with the evaluate classes (module \texttt{evaluate}) which give simple access to many observables calculated over whole trajectories. Alternatively, the data can be returned as NumPy array through the \texttt{Frame} objects and can be analyzed using the individual \amep\ modules, also in combination with self-written Python code to perform additional operations beyond \amep. The \texttt{plot} module allows to visualize analysis results and simulation data in form of figures or videos. Finally, the results can be stored in an HDF5-based data format again (layer ``Result Storage''). It is possible to save and load one or more results in one HDF5 file. Later, the results can be imported as \texttt{DataBase} (for multiple results) or \texttt{EvalData} objects (one result in a file or one result selected from a \texttt{DataBase} object). The red arrows follow the path of our first minimal example as discussed in Section \ref{sec:Usage}. The design itself is explained in more detail in Section \ref{sec:Design}.}
	\label{fig:amep_design}
\end{figure*}

First, we import \amep\ and load the simulation data of non-interacting overdamped ABPs:
\begin{python}
import amep
traj_nonint = amep.load.traj(
    "/path/to/non_interacting_ABPs",
    mode = "lammps"
)
\end{python}
The function \pythoninline{amep.load.traj} creates an HDF5 file (\texttt{traj.h5amep}) in the background that contains all the simulation data and returns a \pythoninline{ParticleTrajectory} object which allows to easily access the data for further processing. Since the data has been produced using LAMMPS \cite{Thompson_CompPhysComm_2022}, we load it using \pythoninline{mode = "lammps"}. To save useful information for long-term storage, we add the author to the trajectory object, which will save this information within the linked HDF5 file:
\begin{python}[firstnumber=6]
traj_nonint.add_author_info(
    "Lukas Hecht",
    "affiliation",
    "Technical University of Darmstadt"
)
traj_nonint.add_author_info(
    "Lukas Hecht",
    "email",
    "lukas.hecht@pkm.tu-darmstadt.de"
)
\end{python}
This information can be accessed by calling \pythoninline{traj_nonint.get_author_info("Lukas Hecht")}, returning
\begin{pythonout}
{'affiliation': 'Technical University of Darmstadt',
 'email': 'lukas.hecht@pkm.tu-darmstadt.de'}
\end{pythonout}
while \pythoninline{traj_nonint.authors} returns a list of all authors. \amep\ provides several methods to add more information to trajectory objects (\pythoninline{ParticleTrajectory} or \pythoninline{FieldTrajectory}) such as software information, simulation scripts, log files, or simulation parameters for a comprehensive and reproducible set of information about the simulation (see online documentation available at \url{https://amepproject.de} and Fig.\ \ref{fig:amep_design}).

Next, we calculate the MSD, which is defined as
\begin{equation}
	\textnormal{MSD}(t) = \frac{1}{N}\sum_{i=1}^{N}\left[\vec{r}_i(t)-\vec{r}_i(0)\right]^2,\label{eq:msd}
\end{equation}
where $\vec{r}_i(0)$, $\vec{r}_i(t)$ denote the position of particle $i$ at time $0$ and $t\geq 0$, respectively. For a single ABP, it can be shown analytically that the MSD can be written as
\begin{align}
	\textnormal{MSD}(t)&= 2l_p^2\left(\dfrac{t}{\tau_p} -1 + e^{-t/\tau_p}\right)+4D_{\rm t}t\nonumber\\
	&\xrightarrow{t\gg \tau_p} 4D_{\rm eff}t \label{eq:msd-abps}
\end{align}
with effective diffusion coefficient $D_{\rm eff}=D_{\rm t}+l_{\rm p}^2/(2\tau_{\rm p})$ \cite{Hecht_ArXiv_2021,Howse_PhysRevLett_2007,TenHagen_JPhysCondensMatter_2011}. Here, $\tau_{\rm p}=1/D_{\rm r}$ denotes the persistence time, which signifies the time after which the directed motion of an ABP is randomized due to rotational diffusion, and $l_{\rm p}=v_0\tau_{\rm p}$ denotes the persistence length, which is the distance an ABP moves on average before its direction of motion is randomized. $D_{\rm t}=k_{\rm B}T_{\rm b}/\gamma_{\rm t}$ and $D_{\rm r}=k_{\rm B}T_{\rm b}/\gamma_{\rm r}$ are the translational and rotational diffusion coefficients, respectively, and $v_0$ is the terminal self-propulsion speed of the ABP. To calculate the MSD with \amep, we create a \pythoninline{MSD} object via
\begin{python}[firstnumber=16]
msd = amep.evaluate.MSD(
    traj_nonint
)
\end{python}
which calculates the MSD for all frames and performs the average over all particles. The \pythoninline{MSD} object contains all information about the MSD and we can access the times via \pythoninline{msd.times} and the value for each individual frame via \pythoninline{msd.frames} for example. The returned objects are NumPy arrays and can therefore easily be used also elsewhere in Python.

To get the effective diffusion coefficient $D_{\rm eff}$ from the MSD in the late time limit, we define the fit function \pythoninline{f} with the fit parameters as keyword arguments and create a corresponding \pythoninline{Fit} object:
\begin{python}[firstnumber=19]
def f(t, D = 1.0):
    return 4*D*t
Dfit = amep.functions.Fit(f)
\end{python}
By calling the object's \pythoninline{fit} method, which uses the \pythoninline{scipy.odr} package in the background, with the data to fit and an initial guess (using the \pythoninline{p0} keyword), we obtain the optimal parameters. Here, we only want to fit the long-time behaviour, and hence, we only consider data for $t>10^1\tau_{\rm p}$. The optimal parameters and their errors can then be retrieved from the \pythoninline{Fit} object:
\begin{python}[firstnumber=22]
mask = msd.times > 1e1
Dfit.fit(
    msd.times[mask],
    msd.frames[mask],
    p0 = [900]
)
print(
    f"D={Dfit.params[0]:0.2f}, "\
    f"D-error={Dfit.errors[0]:0.2f}"
)
\end{python}
\begin{pythonout}
D=885.32, D-error=0.59
\end{pythonout}
This is fairly close to the expected value of $D_{\rm eff}=883$ as obtained from Eq.\ (\ref{eq:msd-abps}) for ABPs with $l_{\rm p}=1.0$, $\tau_{\rm p}=1.0$, $v_0=42$, and $D_{\rm t}=1.0$ as used in this example.

Next, we use the same simulation data to calculate and fit the OACF, which is given by 
\begin{equation}
	\langle \hat{p}(0)\cdot\hat{p}(t) \rangle = \frac{1}{N}\sum_{i=1}^{N}\hat{p}_i(0)\cdot\hat{p}_i(t) \label{eq:oacf}
\end{equation}
and is equal to $e^{-D_{\rm r}t}$ for overdamped ABPs \cite{Sandoval_PhysRevE_2020}. Here, $\hat{p}_i$ is the orientation vector of the effective self-propulsion force of particle $i$. Again, by creating the corresponding evaluate and fit objects, we can analyze the OACF in a few lines of Python code. For the correct normalization, we specify the plane in which the OACF is to be calculated with \pythoninline{direction = "xy"}:
\begin{python}[firstnumber=32]
oacf = amep.evaluate.OACF(
    traj_nonint,
    direction = "xy"
)
def f(t, Dr = 1.0):
    return np.exp(-Dr*t)
Tfit = amep.functions.Fit(f)
Tfit.fit(oacf.times, oacf.frames)
print(
    f"Dr={Tfit.params[0]:0.4f}, "\
    f"Dr-error={Tfit.errors[0]:0.4f}"
)
\end{python}
\begin{pythonout}
Dr=0.9890, Dr-error=0.0021
\end{pythonout}
Again, we obtain a value close to the expected one of $D_{\rm r}=1.0$.

Next, we calculate the RDF. For that, we load simulation data of $N=10^5$ interacting ABPs moving in a two-dimensional periodic simulation box with a total packing fraction of $\varphi=0.5$ and create a \pythoninline{RDF} evaluate object. Note that a corresponding exemplary dataset is available at \url{https://github.com/amepproject/amep/tree/main/examples}. Here, we only want to consider frames that are in the steady state. Therefore, we skip the first 90 \% of frames using \pythoninline{skip = 0.9}. Additionally, we specify the total number of frames to average over (time average) equally spaced in time over the last 90 \% of the trajectory using the \pythoninline{nav} keyword, which is inherited from the \pythoninline{BaseEvaluation} class and is short for \enquote{number of averages}. Due to the large number of particles, it is worth to use multiprocessing, to set a maximum distance until which the RDF is calculated, and to use a large number of bins using the \pythoninline{njobs}, \pythoninline{rmax}, and \pythoninline{nbins} keywords, respectively:
\begin{python}[firstnumber=44]
traj_int = amep.load.traj(
    "/path/to/interacting_ABPs",
    mode = "lammps"
)
rdf = amep.evaluate.RDF(
    traj_int,
    skip = 0.9,
    nav = 100,
    nbins = 20000,
    rmax = 300,
    njobs = 128
)
\end{python}

Before visualizing the results, we save them in HDF5 files. For that, \amep\ offers two possibilities: One can store the result of either a single evaluate object (as done for the RDF) or multiple evaluate objects (as done for the MSD and the OACF) in one HDF5 file:
\begin{python}[firstnumber=56]
msd.save(
    "nonint-results.h5",
    database = True
)
oacf.save(
    "nonint-results.h5",
    database = True
)
rdf.save(
    "rdf.h5"
)
\end{python}
These can later be loaded by calling \pythoninline{msd = amep.load.evaluation("nonint-results.h5", database = True).msd} and \pythoninline{rdf = amep.load.evaluation("rdf.h5")} for example.

\begin{figure*}
	\centering
	\includegraphics[width=0.8\linewidth]{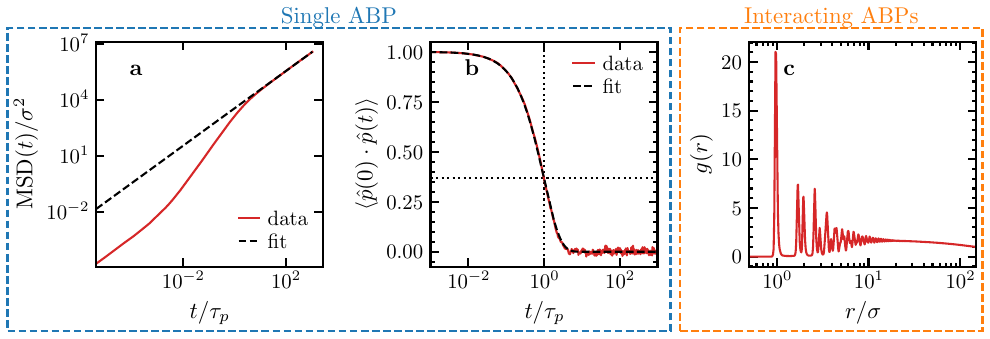}
	\caption{\textbf{How-to-use-\amep\ example.} \textbf{a} Mean-square displacement of a single ABP calculated with \texttt{amep.evaluate.MSD}. The black dashed line is a fit of Eq.\ (\ref{eq:msd-abps}) done with \texttt{amep.functions.Fit}. \textbf{b} Orientation autocorrelation function of a single ABP calculated with \texttt{amep.evaluate.OACF}. The black dashed line is a fit of $e^{-D_{\rm r}t}$ done with \texttt{amep.functions.Fit} and the black dotted lines mark the point $(1,1/e)$. The results in panels a and b are averaged over $4\times 10^3$ particles. \textbf{c} Radial pair distribution function of $N=10^5$ repulsively interacting ABPs calculated using \texttt{amep.evaluate.RDF} and averaged over time in the steady state. The plot has been created using the \texttt{amep.plot} module and the corresponding code is available at \url{https://github.com/amepproject/amep/tree/main/examples}.}
	\label{fig:how_to_use_amep}
\end{figure*}

Finally, the data and fits can be visualized with the Matplotlib wrappers provided by \amep\ (Fig.\ \ref{fig:how_to_use_amep}):
\begin{python}[firstnumber=67]
fig, axs = amep.plot.new(
    (6.5,2), ncols = 3, wspace = 0.1
)
# plot msd
axs[0].plot(
    msd.times, msd.frames,
    "r-", label = "data"
)
axs[0].plot(
    msd.times, Dfit.generate(msd.times),
    "k--", label = "fit"
)
axs[0].set_xlabel("Time")
axs[0].set_ylabel("MSD")
axs[0].loglog()
axs[0].legend()
	
# plot oacf
axs[1].plot(
    oacf.times, oacf.frames,
    "r-", label = "data"
)
axs[1].plot(
    oacf.times, Tfit.generate(oacf.times),
    "k--", label = "fit"
)
axs[1].set_xlabel("Time")
axs[1].set_ylabel("OACF")
axs[1].semilogx()
axs[1].axhline(1/np.e, c = "k")
axs[1].axvline(1, c = "k")
axs[1].legend()
	
# plot rdf
axs[2].plot(rdf.r, rdf.avg, "r-")
axs[2].semilogx()
axs[2].set_xlabel("Distance")
axs[2].set_ylabel("RDF")
	
# plot boxes
amep.plot.draw_box(
    fig, [0, 0, 0.68, 1.01],
    edgecolor = "tab:blue", linestyle = "--",
    text = "Single ABP", c = "tab:blue"
)
amep.plot.draw_box(
    fig, [0.69, 0, 0.31, 1.01],
    edgecolor = "tab:orange", linestyle = "--",
    text = "Interacting ABPs", c = "tab:orange"
)	
# save figure in file
fig.savefig("how_to_use_amep.pdf")
\end{python}
The whole workflow of this minimal example is marked with red arrows in the flowchart shown in Fig.\ \ref{fig:amep_design} and its code is available at \url{https://github.com/amepproject/amep/tree/main/examples}.

% =================================================================================================================
% AMEP DESIGN AND MODULE OVERVIEW
% =================================================================================================================
\section{Design}
\label{sec:Design}
In the previous section, we introduced a short example of how to use \amep. We will now give an overview of the design of the \amep\ package. A more detailed description can be found in the online documentation available at \url{https://amepproject.de}.

% =================================================================================================================
\subsection{Modules}
\label{sub:modules}
\amep\ contains a plethora of pre-implemented functions to analyze particle-based and continuum simulation data. In Tab.\ \ref{tab:modules}, the modules are listed with a short description and a few exemplary functions. The functions contained in these modules take NumPy arrays and native Python data types as input and therefore, can also be used without the data-handling framework provided by \amep. In addition to the modules shown in Tab.\ \ref{tab:modules}, \amep\ has a \pythoninline{plot} module to visualize simulation data and results as well as an \pythoninline{evaluate} module, which provides various evaluation methods that use a full trajectory as input data. The \pythoninline{plot} and the \pythoninline{evaluate} module will be described in more detail below.

% =================================================================================================================
\subsection{Visualization}
\label{sub:visulaization}
For visualizing simulation data and analysis results, \amep\ provides the \pythoninline{plot} module, which is a wrapper for the Matplotlib library \cite{Hunter_SciProg_2007}. The \pythoninline{plot} module includes functions to plot particles and continuum fields (cf.\ Figs.\ \ref{fig:mips} and \ref{fig:fields-snapshots}), to create insets and colorbars (cf.\ Fig.\ \ref{fig:mips}), and to animate plots and simulation data (cf.\ Movies S1--S3, Supplemental Material). For example, a video of a trajectory \pythoninline{traj} can be created via \pythoninline{traj.animate("video.mp4")}. By varying the file extension, also other data formats can be used such as GIFs. Additionally, the \pythoninline{plot} module provides useful functions to format the axis of a figure. All visualization functions are optimized to create figures and videos that can be directly used for a publication in various journals. All figures within this paper with the exception of Fig.\ \ref{fig:amep_design} are created with \amep.

% =================================================================================================================  
\subsection{Evaluate module}
\label{sub:evaluate-objects}
We already introduced the different modules of \amep, which contain a plethora of functions relevant for the analysis of active matter simulation data. The \pythoninline{evaluate} module is a special module in the sense that it uses functions from all other modules to calculate certain observables for a whole trajectory. In our first example in Section \ref{sec:Usage}, we already used the \pythoninline{MSD}, \pythoninline{OACF}, and \pythoninline{RDF} classes, which are part of the \pythoninline{evaluate} module. We will now introduce the concept of these classes, which will be referred to as ``evaluate classes'' in the following. 

While the functions of the modules discussed above take NumPy arrays as input data, evaluate classes take a whole trajectory as an input. One example is the \pythoninline{amep.evaluate.RDF} class used in Section \ref{sec:Usage}. To initialize an object of an evaluate class, the trajectory as well as certain parameters are supplied. The evaluate classes offer the functionality to average over multiple frames automatically (the keyword \pythoninline{nav}, which is an abbreviation for ``number of averages'', allows to specify the number of frames to average over) as well as to skip a certain amount of frames at the beginning of the trajectory (in percent) by specifying the \pythoninline{skip} keyword. The resulting data can be returned as NumPy arrays. Furthermore, one or more evaluate objects can be stored in HDF5 files using their \pythoninline{.save("filename.h5")} method and can be read again with \pythoninline{amep.load.evaluation("filename.h5")}. The evaluate classes are a crucial part in the user-focused design of \amep\ and make it possible to achieve results within a few lines of code.

% =================================================================================================================
\subsection{File format}
\label{sub:h5amep-format}
Based on the Hierarchical Data Format version 5 (HDF5) \cite{HDF5,Collette_Book_PythonAndHDF5_2013}, \amep\ introduces a new file format \texttt{h5amep} to store simulation data and additional metadata. This format is used in the backend of \amep. The HDF5 files are structured into groups and datasets. The \texttt{h5amep} format has the following groups, subgroups, and attributes:
\begin{small}
\begin{verbatim}
h5amep root
\amep
\info
    \authors
    \software
\scripts
\params
\type
\particles or \fields
\frames
    \steps
    \times
    \[frame0]
        \coords
        \velocities
        \...
    \[frame1]
        \...
    \...
\end{verbatim}
\end{small}
The group \texttt{amep} contains information about the \amep\ version that has been used to create the \texttt{h5amep} file. The group \texttt{info} contains the saved information about authors (cf.\ Section \ref{sec:Usage}) and software. The \texttt{scripts} group gives the possibility to save text files such as simulation scripts and log files that correspond to the simulation data. In the \texttt{params} group, \amep\ stores parameters such as the simulation timestep for example. Additional simulation parameters can be added. The attribute \texttt{type} contains a flag about the type of data stored in the \texttt{h5amep} file. This can be either \pythoninline{"particle"} or \pythoninline{"field"}. The groups \texttt{particles} and \texttt{fields} contain user-defined information about the particles and the fields used in the simulation, respectively. Finally, the group \texttt{frames} contains multiple datasets and subgroups with the simulation data. The dataset \texttt{steps} stores a list of all the frame numbers (i.e., the number of timesteps for each frame) and the dataset \texttt{times} the corresponding (physical) time, while the individual frames of the simulation are stored in subgroups of \texttt{frames} named by their simulation step. Within an individual frame, the simulation data is stored, e.g., coordinates and velocities for particle-based simulations or density and concentrations for continuum simulations, as separate datasets.

\begin{table}
	\centering
	\caption{\textbf{Module overview.} The shown \amep\ modules take NumPy arrays and native Python data types as input. The description of each module is complemented with a few exemplary functions (italic text). \amep\ additionally provides the modules \texttt{plot} and \texttt{evaluate} as described in Section \ref{sec:Design}. A detailed description of all modules and the contained functions can be found in the \amep\ online documentation available at \url{https://amepproject.de}.}
	\small{
		\begin{tabular}{l|p{6cm}}
			\textbf{Module:} & \textbf{Description:}\\
			\hhline{==}
			\texttt{cluster} & cluster identification and properties \\
			& \textit{identify clusters (particles)} \\
			& \textit{radius of gyration} \\
			% & \textit{centre of mass} \\
			\hline
			\texttt{continuum} & field data analysis and coarse-graining \\
			& \textit{identify clusters (fields)} \\
			& \textit{create field from particles} \\
			% & \textit{2d spatial correlation function} \\
			\hline
			\texttt{functions} & mathematical functions and fitting\\
			& \textit{Gaussian} \\
			& \textit{general fit class} \\
			% & \textit{maxwell-boltzmann} \\
			\hline
			\texttt{order} & functions to characterize spatial order \\
			& \textit{local density} \\
			& \textit{k-atic bond order parameter} \\
			% & \textit{number of next neighbours} \\
			\hline
			\texttt{pbc} & handling of periodic boundary conditions \\
			& \textit{create periodic images} \\
			& \textit{fold coordinates back into the box} \\
			% & \textit{} \\
			\hline
			\texttt{spatialcor} & spatial correlation functions \\
			& \textit{radial distribution function} \\
			& \textit{isotropic structure factor} \\
			% & \textit{1d, 2d spatial correlation functions} \\
			\hline
			\texttt{statistics} & methods for statistical analysis \\
			& \textit{binder cumulant} \\
			& \textit{distribution (histograms) 1d, 2d} \\
			% & \textit{2d histogram} \\
			\hline
			\texttt{thermo} & thermodynamic observables \\
			& \textit{kinetic temperature} \\
			& \textit{kinetic energies} \\
			% & \textit{} \\
			\hline
			\texttt{timecor} & time correlation functions \\
			& \textit{mean squared distance} \\
			& \textit{autocorrelation function} \\
			% & \textit{incoherent intermediate scattering function} \\
			\hline
			\texttt{utils} & utility methods \\
			& \textit{averaging methods (running mean, ...)} \\
			& \textit{Fourier transform} \\
			% & \textit{check if in box} \\
			% \hline
			% \texttt{amep.evaluate} & evaluation classes for analysis of full trajectories (cf. \autoref{sec:evaluate_objects}) \\
			% & \textit{MSD} \\
			% & \textit{RDF} \\
			% % & \textit{PCFangle} \\
			\bottomrule
		\end{tabular}
	}
	\label{tab:modules}
\end{table}

% =================================================================================================================
\subsection{Trajectories and frames}
\label{sub:traj-frames}
The \texttt{h5amep} file is connected to Python via the \pythoninline{amep.load} module with which the simulation data is imported to a \pythoninline{ParticleTrajectory} or \pythoninline{FieldTrajectory} object. Saved evaluation objects can also be imported with the \pythoninline{amep.load} module. Because a \pythoninline{ParticleTrajectory} or \pythoninline{FieldTrajectory} object works as a reader for the \texttt{h5amep} file, the simulation data (coordinates, velocities, density, concentration,...) is not stored in the main memory all the time but only the portion that is requested for the specific analysis. The simulation data can be accessed via \pythoninline{BaseFrame} or \pythoninline{BaseField} objects for particle-based and continuum data, respectively, which will be referred to as ``frame'' in the following. Technically, the \pythoninline{ParticleTrajectory} or \pythoninline{FieldTrajectory} object acts as a list of frames, i.e., \pythoninline{frame = traj[0]} returns the first frame of the trajectory \pythoninline{traj}. The data of one frame can be accessed through various methods. For example, the coordinates and velocities of particles can be accessed via \pythoninline{frame.coords()} and \pythoninline{frame.velocities()}. Other data can be accessed via the \pythoninline{frame.data} method, e.g., \pythoninline{frame.data("mass")} returns the mass of each particle and \pythoninline{frame.data("rho")} the density field of a continuum simulation. A list of all available keys can be accessed via \pythoninline{frame.keys}. All these methods return NumPy arrays for efficient calculations and simple integrity to other Python packages (see also Fig.\ \ref{fig:amep_design}).

% =================================================================================================================
% PARTICLES
% =================================================================================================================
\section{Analyzing particle-based simulation data with \amep}
\label{sec:Particles}
\begin{figure*}
    \centering
    \includegraphics[width=1.0\linewidth]{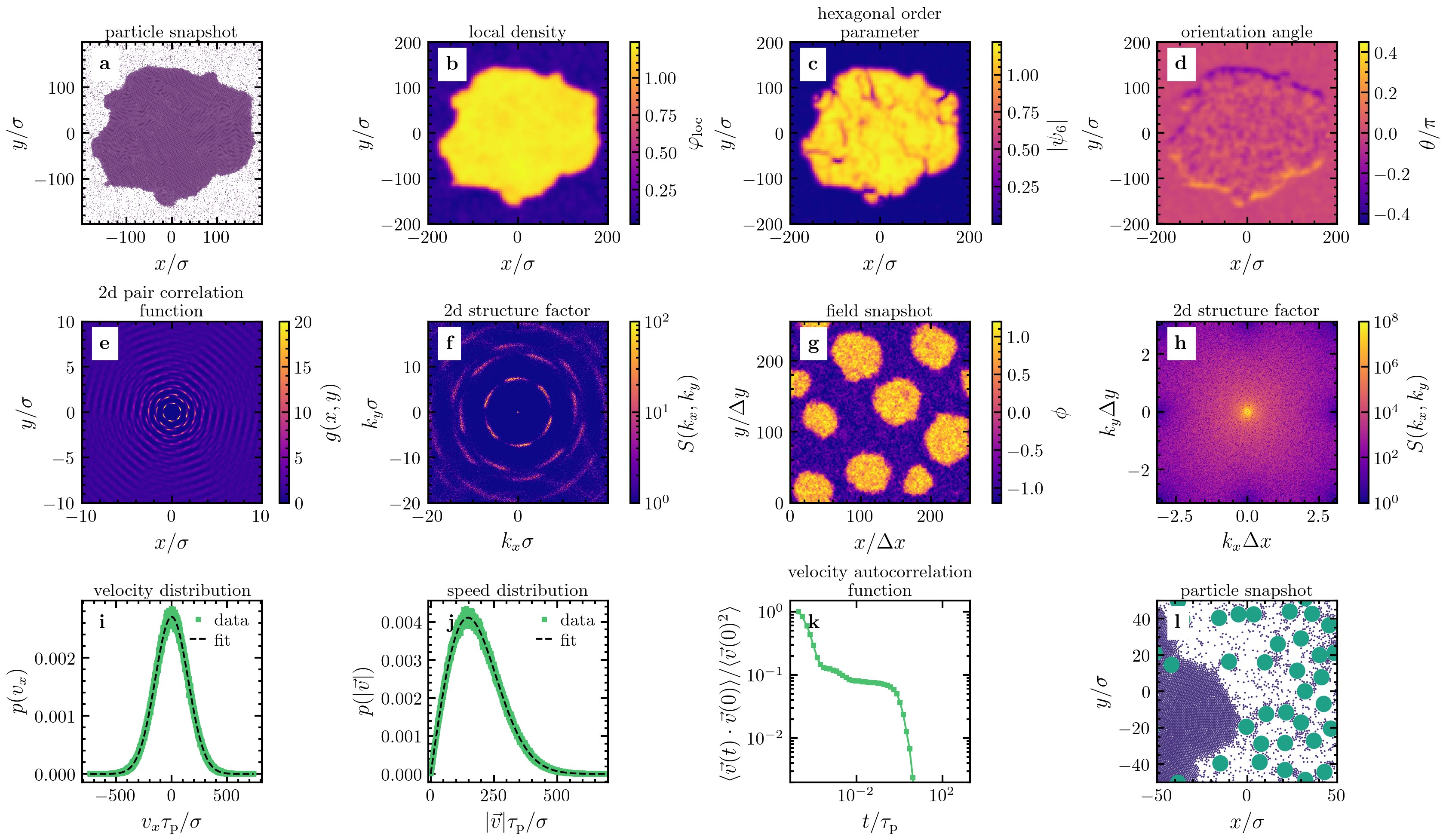}
    \caption{\textbf{Further observables calculated with \amep:} \textbf{a} Snapshot of MIPS plotted with \texttt{amep.plot.particles}. \textbf{b--d} Corresponding coarse-grained local density $\varphi_{\rm loc}$, hexagonal order parameter $\psi_6$ [Eq.\ (\ref{eq:psi6})], and orientation angle $\theta$, respectively, calculated with \texttt{amep.continuum.gkde}. \textbf{e} Two-dimensional pair correlation function of the snapshot shown in panel a calculated with \texttt{amep.evaluate.PCF2d}. \textbf{f} Two-dimensional structure factor of the snapshot shown in panel a calculated with \texttt{amep.evaluate.SF2d}. \textbf{g} Snapshot of MIPS in the active model B+ plotted with \texttt{amep.plot.field}. \textbf{h} Two-dimensional structure factor of the field shown in panel g calculated with \texttt{amep.evaluate.SF2d}. \textbf{i,j} Distribution of the velocity $v_x$ in $x$ direction and the magnitude $|\vec{v}|$ from a particle-based simulation in two spatial dimensions calculated with \texttt{amep.evaluate.VelDist}. The black dashed lines are fits obtained with \texttt{amep.functions.NormalizedGaussian} and \texttt{amep.functions.MaxwellBoltzmann}, respectively. \textbf{k} Velocity autocorrelation function calculated with \texttt{amep.evaluate.VACF}. \textbf{l} Snapshot of a simulation with particles of different sizes plotted with \texttt{amep.plot.particles}.}
    \label{fig:overview}
\end{figure*}

We will now use \amep\ to analyze simulation data of large-scale simulations of ABPs. After introducing the simulation details, we demonstrate how to calculate certain observables using \amep\ and briefly discuss the results. The code examples demonstrate the workflow of \amep\ and serve as a guide for using \amep. Further observables that are available in \amep\ but which are not discussed in this section are exemplarily shown in Fig.\ \ref{fig:overview}.

\subsection{Brownian dynamics simulations of active Brownian particles}
\label{sub:simulation-details}
As a first detailed example, we analyze particle-based simulation data obtained with the active Brownian particle (ABP) model as introduced in Section \ref{sec:Usage} as Eqs.\ (\ref{eq:abps-trans}) and (\ref{eq:abps-rot}). The repulsive interaction is modeled by the Weeks-Chandler-Anderson (WCA) potential $u(r_{ji})=4\epsilon[(\sigma/r_{ji})^{12} -(\sigma/r_{ji})^{6}]+\epsilon$, where $r_{ji}=\left\lvert\vec{r}_j-\vec{r}_i\right\rvert$ and $u=0$ for $r_{ji}/\sigma \ge 2^{1/6}$ \cite{Weeks_JCP_1971}. For the simulations here, we have chosen $l_{\rm p}=100\sigma$, $\epsilon/k_{\rm B}T_{\rm b}=1.0$, and for simplicity, $\gamma_{\rm t}=\gamma_{\rm r}/\sigma^2$ unless otherwise stated (note that in experiments of active granulates, the Stokes-Einstein relation does not apply \cite{Scholz_NatCom_2018}). Here, $D_{\rm t}=k_{\rm B}T_{\rm b}/\gamma_{\rm t}$ and $D_{\rm r}=k_{\rm B}T_{\rm b}/\gamma_{\rm r}$ are the translational and rotational diffusion coefficients, respectively. We define the P\'eclet number as ${\rm Pe}=v_0/\sqrt{2D_{\rm t}D_{\rm r}}$, which quantifies the strength of self-propulsion relative to diffusive motion. Note that all simulations are done in the overdamped regime, i.e., we choose a small mass $m/(\gamma_{\rm t}\tau_{\rm p})=5\times 10^{-5}$. Since it has been found in previous works that the influence of the moment of inertia $I$ on the simulation results is rather unimportant \cite{Mandal_PhysRevLett_2019}, it is held constant at $I/(\gamma_{\rm r}\tau_{\rm p})=0.33$ unless otherwise stated. All simulations are done within a two-dimensional quadratic box of length $L$ with periodic boundary conditions and up to $N=1.28 \times 10^6$ particles with a time step of $\Delta t/\tau_{\rm p}=10^{-5}$--$10^{-6}$ using LAMMPS \cite{Thompson_CompPhysComm_2022}. We choose ${\rm Pe}=70.7$ and a packing fraction of $\varphi=0.5$ with $\varphi=N\pi\sigma^2/(4L^2)$ and start each simulation from uniformly distributed particle positions unless otherwise stated.

\amep\ can load the common LAMMPS plain text format in which LAMMPS stores one snapshot of the simulation per text file named as \texttt{dump*.txt} for example, where the \texttt{*} is a placeholder for the current time step. These files are formatted as follows:
\begin{scriptsize}
\begin{verbatim}
ITEM: TIMESTEP
20000
ITEM: NUMBER OF ATOMS
1280000
ITEM: BOX BOUNDS pp pp pp
0.0000000000000000e+00 1.5000000000000000e+03
0.0000000000000000e+00 1.5000000000000000e+03
-5.0000000000000000e-01 5.0000000000000000e-01
ITEM: ATOMS id type x y z fx fy mux muy radius mass
1 1 26.206 44.939 0 95.5607 9.87064 0.99913 0.04168 0.5 0.00005
2 1 45.187 24.985 0 -94.141 -33.726 -0.9414 -0.3373 0.5 0.00005
3 1 53.126 32.953 0 -86.024 12.4595 -0.9831 0.18303 0.5 0.00005
...
\end{verbatim}
\end{scriptsize}
Here, \texttt{id} denotes a uniqe identifier for each particle, \texttt{type} denotes the particle type, \texttt{x,y,z} denote the position of each particle, \texttt{fx, fy} denote the forces acting on each particle, \texttt{mux, muy} are the components of the orientation vector $\hat{p}$, \texttt{radius} denotes the radius $\sigma/2$ of each particle, and \texttt{mass} their mass $m$. Note that \texttt{z} is zero for all particles because the simulation is done in two spatial dimensions. An exemplary dataset is available at \url{https://github.com/amepproject/amep/tree/main/examples}. \amep\ reads all data from each text file and converts it into the \texttt{h5amep} format. Each item from the \texttt{ATOMS} section in the LAMMPS text file can then be accessed with the \pythoninline{.data()} method of the corresponding \pythoninline{BaseFrame} object, e.g., the force in $x$ direction of frame 5 can be returned as NumPy array by calling 
\begin{python}
import amep
traj = amep.load.traj(
    "/path/to/data",
    mode = "lammps"
)
fx = traj[5].data("fx")
\end{python}
For many standard quantities such as coordinates, forces, or velocities, \amep\ provides further commands exemplarily shown below:\footnote{See online documentation at \url{https://amepproject.de} for further details.}
\begin{python}[firstnumber=7]
coords = traj[5].coords()
forces = traj[5].forces()
mu = traj[5].data("mux", "muy")
\end{python}

\begin{figure*}
	\centering
	\includegraphics[width=0.75\linewidth]{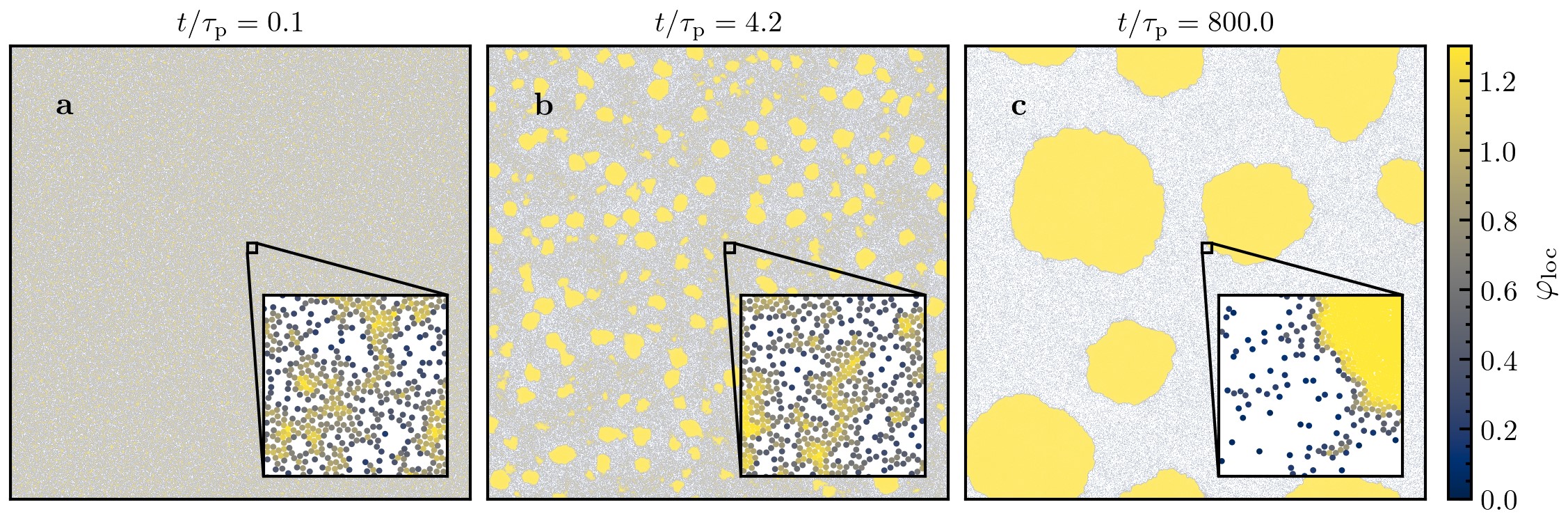}
	\caption{\textbf{Plotting snapshots with \texttt{amep.plot.particles}:} \textbf{a--c} Snapshots of ABPs undergoing MIPS starting from a uniform distribution for three different times as given in the key. The particles are colored with respect to their local area fraction $\varphi_{\rm loc}$ as obtained from a Voronoi tesselation calculated using \texttt{amep.order.voronoi\_density} (see Subsection \ref{sub:particles-local-order} for further details). The insets show the indicated extract of the simulation box and have been plotted using \texttt{amep.plot.add\_inset}. Parameters: $m/(\gamma_{\rm t}\tau_{\rm p})=5\times 10^{-5}$, $I/(\gamma_{\rm r}\tau_{\rm p})=0.33$, $N=1.28\times 10^6$, $\varphi=0.5$, ${\rm Pe}=70.7$.}
	\label{fig:mips}
\end{figure*}

\subsection{Motility-induced phase separation}
\label{sub:particles-mips}
For large enough packing fraction and large enough Pe, ABPs can phase separate into a dense and a dilute phase. This phenomenon is well-known as motility-induced phase separation (MIPS) \cite{Digregorio_PhysRevLett_2018,DeKarmakar_SoftMatter_2022,Redner_PhysRevLett_2013,Cates_AnnuRevCondensMatterPhys_2015,Su_CommPhys_2023,Redner_PhysRevLett_2016,Takatori_PhysRevE_2015,Speck_PhysRevLett_2014,Tailleur_PhysRevLett_2008,Mandal_PhysRevLett_2019,Su_CellReportsPhysicalScience_2024} and can be understood as follows: If two ABPs collide, they can block each other due to their effective self-propulsion force. The collision can be resolved if the ABPs reorient their self-propulsion direction due to rotational diffusion and the self-propulsion direction of one ABP deviates from the other. Small clusters can form if more ABPs collide with the two ABPs blocking each other. Roughly, if now the mean time $\tau_{\rm c}$ between collisions is smaller than the mean time $\tau_{\rm p}$ a particle randomly reorients its self-propulsion direction, small clusters tend to grow, which finally results in a phase separation where an active gas coexists with a dense liquid \cite{Farrell_PhysRevLett_2012,Cates_EPL_2013,Cates_AnnuRevCondensMatterPhys_2015,Gonnella_CRPhysique_2015}. This process is exemplarily visualized by the snapshots shown in Fig.\ \ref{fig:mips}, which have been created using the \pythoninline{amep.plot.particles} function. In the following, we will mainly focus on the analysis of the simulation visualized in Fig.\ \ref{fig:mips}.

\subsection{Voronoi diagrams and local order}
\label{sub:particles-local-order}
Whether a simulation undergoes MIPS can be quantified by calculating the distribution of the local density or local area fraction, which is unimodal in the uniform regime and becomes bimodal in the MIPS regime \cite{Digregorio_PhysRevLett_2018,Su_NewJPhys_2021,Klamser_NatCom_2018,DeKarmakar_SoftMatter_2022}. \amep\ provides multiple functions to calculate the local density or area fraction: \pythoninline{amep.order.local_number_density}, \pythoninline{amep.order.local_mass_density}, and \pythoninline{amep.order.local_packing_fraction} determine the local number density, mass density, and area/volume fraction, respectively, from averages over circles of radius $R$, which can be written as $\varphi_{\rm loc}(\vec{r}_i)=\sum_{j=1}^{N}\sigma_j^2{\rm H}(R-|\vec{r}_i-\vec{r}_j|)/(4 R^2)$ in case of the local area fraction. Here, H is the Heaviside step function and $\sigma_j$ is the diameter of particle $j$. Alternatively, \pythoninline{amep.order.voronoi_density} calculates the local mass density, number density, or area/volume fraction based on a Voronoi tesselation. Here, we will use the latter for calculating the distribution of (i) the local area fraction and (ii) the number of next neighbors. 

The Voronoi diagram serves as a method for modeling the structures of materials across various disciplines, including crystallography, ecology, astronomy, epidemiology, geophysics, computer graphics, and more \cite{Stewart_EcoMod_2010,Arcelli_PatRecLett_1986,Chen_FrontGen_2017,Vavilova_InBook_VoronoiTesselation_2021,Hu_ComputGeo_2014}. In essence, when given a set of points in the plane, the Voronoi diagram divides the plane based on the nearest neighbor rule, associating each point with the region of the plane closest to it. The mathematical definition extends to $N$-dimensions, often referred to as Voronoi tessellation \cite{Aurenhammer_ACMComputSurv_1991,Burns_Article_Voronoi}. For a set of finite points $P = \{p_1,p_2, \cdots, p_n \}$ in a subspace $\zeta$ in the Euclidean space, the Voronoi cell is formed through the division of the plane into regions, where each region encompasses all points that are closer to each $p_i$. Points equidistant from at least two points in set $P$ are not part of the Voronoi cell. Instead, they constitute the Voronoi edges. The compilation of all Voronoi cells forms the Voronoi tessellation associated with $\zeta$.

\begin{figure*}
	\centering
	\includegraphics[width=0.75\linewidth]{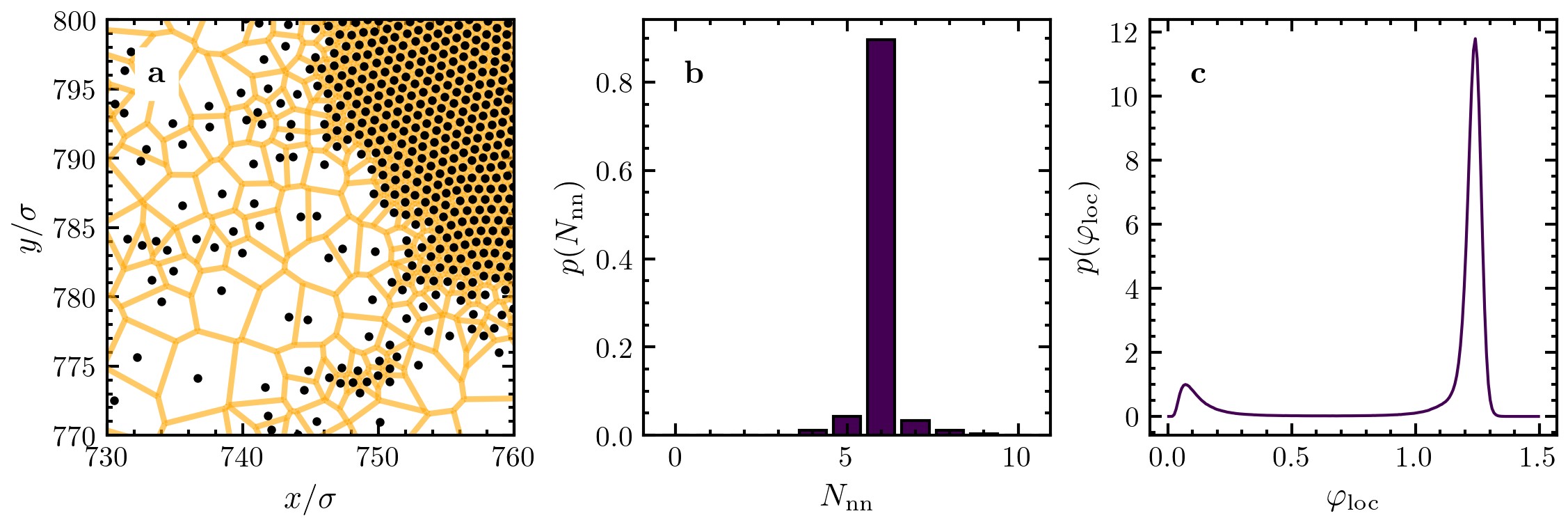}
	\caption{\textbf{Analyzing local order with the \texttt{amep.order} and \texttt{amep.evaluate} modules:} \textbf{a} Same extract of the simulation box as shown in the inset of Fig.\ \ref{fig:mips}c now with the corresponding Voronoi diagram obtained from \texttt{amep.order.voronoi} and plotted with \texttt{amep.plot.voronoi}. The black circles denote the particle positions, the orange lines the border of each Voronoi cell. \textbf{b} Distribution of the number of next neighbors obtained from the Voronoi diagram using \texttt{amep.order.next\_neighbors} and \texttt{amep.statistics.distribution}. \textbf{c} Distribution of the local area fraction obtained from the areas of the Voronoi cells using \texttt{amep.evaluate.LDdist}. The results in panels b and c are averaged over five independent ensembles as exemplary demonstrated in the code examples below. Parameters are the same as in Fig.\ \ref{fig:mips}.}
	\label{fig:voronoi}
\end{figure*}

The Voronoi tesselation is included in the \pythoninline{order} module of \amep\ as function \pythoninline{amep.order.voronoi}. It calculates the Voronoi diagram for a given set of particle coordinates using the \pythoninline{scipy.spatial.Voronoi} class of the SciPy package \cite{Virtanen_NatMeth_2020}. Here, we use the Voronoi tesselation to calculate the number of next neighbors and the local area fraction. Let us first calculate the Voronoi diagram of the last frame (\pythoninline{traj[-1]}) via
\begin{python}
import amep
# load the simulation data
traj = amep.load.traj(
    "/path/to/data",
    mode = "lammps"
)
# calculate the Voronoi diagram
vor, ids = amep.order.voronoi(
    traj[-1].coords(),
    traj[-1].box,
    pbc = True
)
\end{python}
The obtained Voronoi diagram is demonstrated in Fig.\ \ref{fig:voronoi}a for the simulation snapshot shown in Fig.\ \ref{fig:mips}c. It can be easily visualized using \pythoninline{amep.plot.voronoi}. Next, we calculate the number of next neighbors from the Voronoi diagram and its distribution via
\begin{python}[firstnumber=13]
# number of next neighbors for each particle
nnn, _, _, _, _ = amep.order.next_neighbors(
    traj[-1].coords(),
    traj[-1].box,
    vor = vor,
    ids = ids,
    pbc = True
)
# distribution of the number of next neighbors
nndist, bins = amep.statistics.distribution(
    nnn
)
\end{python}
As we can see in Fig.\ \ref{fig:voronoi}b, the distribution of the number of next neighbors attains its highest point at $N_{\rm nn}=6$. This observation indicates the existence of a hexagonal structure within the dense phase, as we will discuss in more detail below. Finally, we calculate the local area fraction using the Voronoi areas and evaluate its distribution:
\begin{python}[firstnumber=25]
# local area fraction of each particle
ld = amep.order.voronoi_density(
    traj[-1].coords(),
    traj[-1].box,
    radius = traj[-1].radius(),
    mode = "packing",
    vor = vor,
    ids = ids
)
# distribution of the local area fraction
lddist, bins = amep.statistics.distribution(
    ld
)
\end{python}
If one would like to calculate the distribution of the local area fraction averaged over several frames of a trajectory, i.e., using a time average, one can simply create an \pythoninline{amep.evaluate.LDdist} object which is directly performing the time average and allows us to store the results in an HDF5 file. Additionally, multiple results can be stored in the same HDF5 file using \amep's database feature: As an example, we calculate the distribution of the local area fraction for five independent simulation runs (ensembles) stored in directories \texttt{/path/to/data/do00}--\texttt{/path/to/data/do04} and store them together in one HDF5 file \texttt{ld.h5}:
\begin{python}
import amep
# load all trajectories
trjs = [
    amep.load.traj(
        f"/path/to/data/do{i:02}",
        mode = "lammps"
    ) for i in range(5)
]
for i,traj in enumerate(trjs):
    # distribution of the local area fraction
    ld = amep.evaluate.LDdist(
        traj,
        nav = traj.nframes,
        use_voronoi = True,
        mode = "packing"
    )
    # set name under which the data
    # is stored in the HDF5 file
    ld.name = f"do{i:02}"

    # save all in file ld.h5
    ld.save("ld.h5", database = True)
\end{python}
Now, we can calculate the ensemble average:
\begin{python}[firstnumber=23]
# load the analysis results
lds = amep.load.evaluation(
    "ld.h5", database = True
)
# ensemble average
ensavg = 0.0
for i in range(5):
    ensavg += lds[f"do{i:02}"].avg/5
\end{python}
The ensemble-averaged result is shown in Fig.\ \ref{fig:voronoi}c. As expected, the distribution of the local area fraction shows two peaks indicating that the system is phase-separated into a dense and a dilute phase. Note that the high-density peak is located at $\varphi_{\rm loc}>1.0$ due to the softness of the particles.

\subsection{Hexagonal order}
\label{sub:particles-hexorder}
\begin{figure*}
	\centering
	\includegraphics[width=0.75\linewidth]{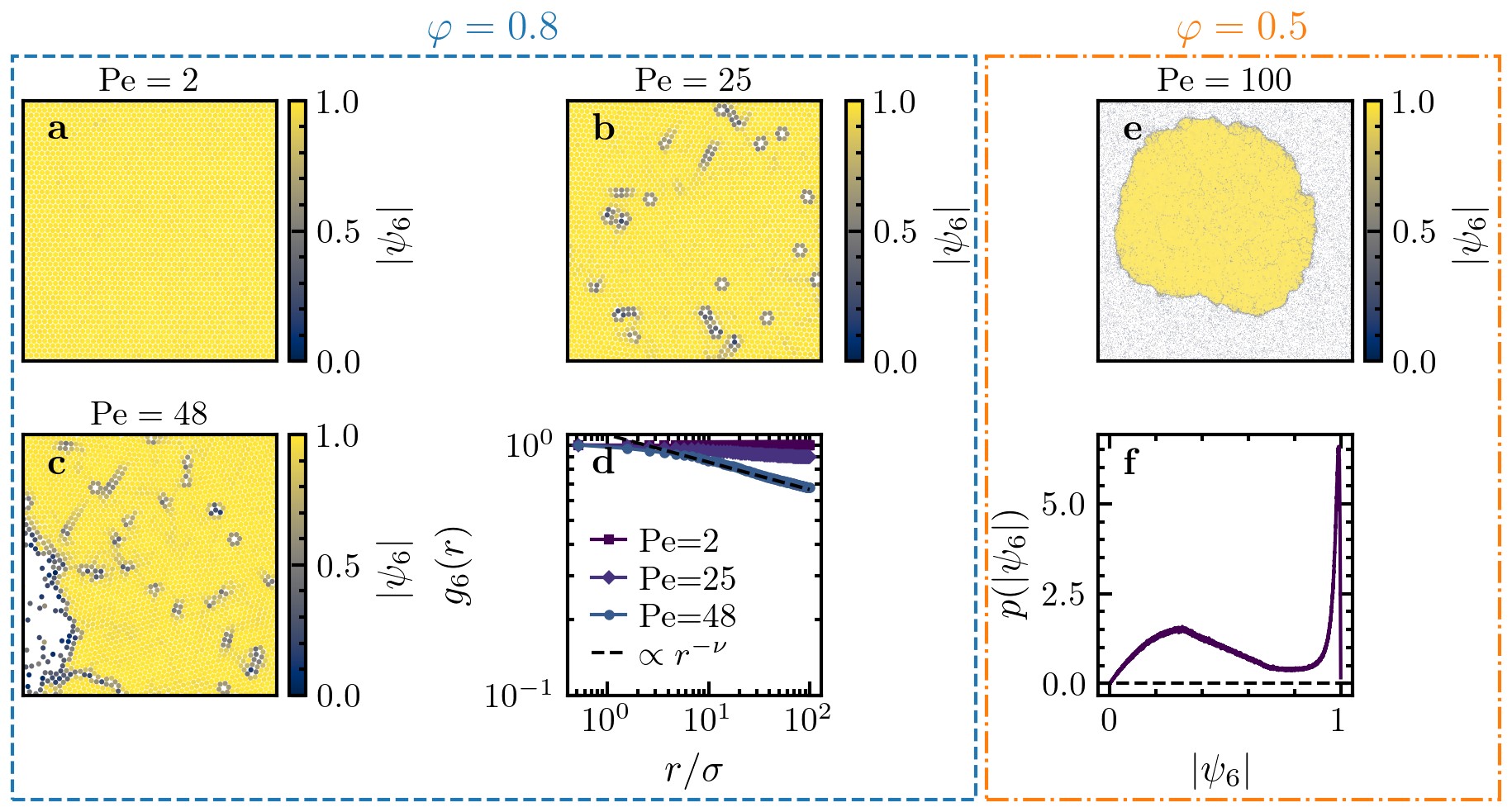}
	\caption{\textbf{Analyzing the hexagonal order using the \texttt{amep.order}, \texttt{amep.spatialcor}, and \texttt{amep.evaluate} modules:} \textbf{a--c} Extracts from the simulation box of an ABP system at high area fraction $\varphi=0.8$ for ${\rm Pe}=2,25,48$, respectively. The particles are colored with respect to their hexagonal order parameter defined in Eq.\ (\ref{eq:psi6}) and calculated with \texttt{amep.order.psi\_k}. Here, the simulations started from a hexagonal crystal as initial condition. \textbf{d} Spatial correlation function $g_6(r)$ as defined in Eq.\ (\ref{eq:psi6-cor}) for the three snapshots shown in panels a--c averaged over three independent ensembles and over time in the steady state using \texttt{amep.evaluate.HexOrderCor}. The black dashed line shows an algebraic decay of the form $r^{-\nu}$ with exponent $\nu=0.11$ as a guide for the eye. \textbf{e} Snapshot of one cluster of the simulation shown in Fig.\ \ref{fig:mips}c with ${\rm Pe}=70.7$ and $\varphi=0.5$. Again, the particles are colored with respect to their hexagonal order parameter. \textbf{f} Distribution of the hexagonal order parameter corresponding to the snapshot shown in panel e calculated with \texttt{amep.evaluate.Psi6dist} and averaged over five independent ensembles. Parameters: $I/(\gamma_{\rm r}\tau_{\rm p})=5\times 10^{-6}$, $N=100\,000$, $\varphi=0.8$, and $\epsilon/(k_{\rm B}T_{\rm b})=10$ (a--d), and (e--f) as in Fig.\ \ref{fig:mips}.}
	\label{fig:hexorder}
\end{figure*}

In the last paragraph, we saw that on average, each particle has six next neighbors, indicating that a hexagonal order dominates the local order of the particles. To quantify the local order in more detail, we calculate the hexagonal order parameter
\begin{equation}
	\psi_{6}(\vec{r}_i)=\frac{1}{6}\sum_{j=1}^{6}\exp\left\lbrace {\rm i} 6\theta_{ij}\right\rbrace
	\label{eq:psi6}
\end{equation}
of each particle, where the sum goes over the six nearest neighbors of particle $i$ and $\theta_{ij}$ is the angle between the connection line from $\vec{r}_i$ to $\vec{r}_j$ and the $x$ axis \cite{Digregorio_PhysRevLett_2018,Nelson_PhilMagA_1982,Cugliandolo_ArXiv_2018}. In two spatial dimensions, the particles within the dense phase typically show local hexagonal order while they are disordered in the dilute phase. Therefore, the distribution of $\psi_6$ can also be used to assess whether a system is phase separated similarly to the local area fraction \cite{Digregorio_PhysRevLett_2018,Cugliandolo_ArXiv_2018,Paoluzzi_CommPhys_2022,Bernard_PhysRevLett_2011}. Calculating the distribution of the hexagonal order parameter can be easily done using \amep's evaluate module:
\begin{python}
import amep
# load the simulation data
traj = amep.load.traj(
    "/path/to/data",
    mode = "lammps"
)
# psi6 distribution (all frames)
p6dist = amep.evaluate.Psi6dist(
    traj, 
    nav = traj.nframes
)
# save results in HDF5 file
p6dist.save("p6dist.h5")

# plot the result of the last frame
fig, ax = amep.plot.new()
ax.plot(
    p6dist.psi6,
    p6dist.frames[-1,0]
)
fig.savefig("p6dist.png")
\end{python}
The result is demonstrated in Fig.\ \ref{fig:hexorder}f, where we have averaged over five independent ensembles. The broad peak at small values corresponds to the dilute phase while the sharp peak at $|\psi_6|=1$ corresponds to the hexagonally packed dense phase as shown in Fig.\ \ref{fig:hexorder}e. It is also common to analyze the spatial range of the hexagonal order to check whether there exists a short-range or (quasi) long-range hexagonal order like in a crystal. To this end, one can calculate the spatial correlation function of the hexagonal order parameter
\begin{equation}
	g_6(r) = \frac{\left\langle\psi_6^{*}(\vec{r}_i)\psi_6(\vec{r}_j)\right\rangle}{\left\langle|\psi_6(\vec{r}_i)|^2\right\rangle}
	\label{eq:psi6-cor}
\end{equation}
with $r=|\vec{r}_i-\vec{r}_j|$ \cite{Digregorio_PhysRevLett_2018,Cugliandolo_ArXiv_2018}. Here, we calculate $g_6(r)$ for ABP simulations  with $\epsilon/(k_{\rm B}T_{\rm b})=10$, $N=100\,000$ particles, and at $\varphi=0.8$ starting from a hexagonal lattice instead of a uniform distribution and average over multiple frames in the steady state:
\begin{python}
import amep
# load simulation data
traj = amep.load.traj(
    "/path/to/data",
    mode = "lammps"
)
# calculate g6
g6 = amep.evaluate.HexOrderCor(
    traj,
    nav = traj.nframes,
    njobs = 16,
    rmax = 100.0,
    skip = 0.5
)
# save results in an HDF5 file
g6.save(
    "g6.h5"
)
# plot average
fig, ax = amep.plot.new()
ax.plot(
    g6.r,
    g6.avg
)
fig.savefig("g6.png")
\end{python}
Since this calculation is computationally expensive, we speed it up using parallelization by specifying the number of jobs that should run in parallel with the \pythoninline{njobs} keyword. Additionally, we give a maximum distance up to which the correlation function should be calculated and we skip the first 50\% of the trajectory to ensure that only the second half of the trajectory, which is in the steady state, is used for the calculation. The result is shown in Fig.\ \ref{fig:hexorder}d together with the corresponding snapshots in Fig.\ \ref{fig:hexorder}a--c for three different Pe. In accordance to Ref.\ \cite{Digregorio_PhysRevLett_2018}, we observe a constant $g_6$ at small Pe and an algebraic decay of the form $r^{-\nu}$ with exponent $\nu$ at higher Pe.

\subsection{Structure factor and coarsening}
\label{sub:particles-sf}
To further probe the order of an active system, one can exploit the radial distribution function $g(r)$, which we have exemplarily calculated in Section \ref{sec:Usage} using \pythoninline{amep.evaluate.RDF}. From an experimental viewpoint, it is easier to obtain the structure factor $S(\vec{q},t)$, which is defined as \cite{Hansen_Book_TheoryOfSimpleLiquids_2006}
\begin{equation}
	S(\vec{q},t)=\frac{1}{N}\left\langle \sum_{i=1}^{N} \sum_{j=1}^{N} \exp\left\lbrace{\rm i}\vec{q}\cdot[\vec{r}_i(t) - \vec{r}_j(t)] \right\rbrace\right\rangle.
	\label{eq:structurefactor}
\end{equation}
Here, $\vec{q}$ represents the wave vector, $\vec{r}_i(t)$ and $\vec{r}_j(t)$ denote the positions of particles $i$ and $j$ at time $t$, respectively, and $N$ is the total number of particles. For isotropic systems, the structure factor only depends on the magnitude $q=|\vec{q}|$ of the wave vector. It gives information about the density response to an external perturbation of wave length $2\pi/q$ and can be probed experimentally via scattering experiments. Note that $S(\vec{q},t)$ is directly related to  $g(\vec{r},t)$ via Fourier transform. However, in practice, it is often preferable to calculate $S(\vec{q}, t)$ directly to achieve good numerical results in the low $q$ regime. Within \amep, $S(\vec{q},t)$ can be directly calculated with \pythoninline{amep.evaluate.SF2d} and the isotropic structure factor $S(q,t)$ with $q=|\vec{q}|$ (i.e., $S(\vec{q},t)$ averaged over the direction of $\vec{q}$) with \pythoninline{amep.evaluate.SFiso}. Here, we use $S(q,t)$ to analyze the coarsening process of the clusters forming when the ABPs undergo MIPS. To this end, we deduce a characteristic length scale $L(t)$ based on the first moment of $S(q,t)$ given by \cite{Stenhammar_PhysRevLett_2013,Stenhammar_SoftMatter_2014,Wittkowski_NatComm_2014,Kendon_JGluidMech_2001,Zhang_SoftMatter_2019}
\begin{equation}
	L(t) =\frac{ \int_{2\pi/L}^{q_\text{cut}} S(q,t) \diff q}{\int_{2\pi/L}^{q_\text{cut}} q S(q,t) \diff q},
	\label{eq:Lt}
\end{equation}
where $L$ is the length of the simulation box. We set the upper limit $q_{\rm cut} \sigma=0.3$, i.e., we only consider length scales larger than $2\pi/q_{\rm cut}\approx 21\sigma$. Notably, under suitable conditions, overdamped ABPs undergoing MIPS exhibit an effective mapping onto a suitable equilibrium system at coarse-grained scales, as established in the literature \cite{Speck_PhysRevLett_2014}. This mapping explains why the coarsening dynamics follows the universal law $L(t) \sim t^{1/3}$ -- a characteristic scaling behavior observed in equilibrium systems.

\begin{figure}
	\centering
	\includegraphics[width=1.0\linewidth]{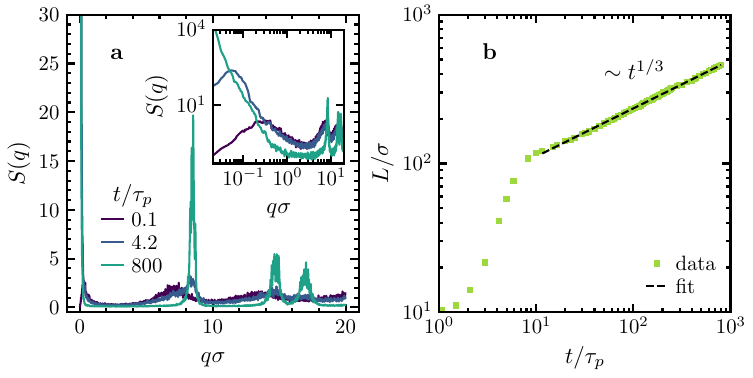}
	\caption{\textbf{Structure factor from \texttt{amep.evaluate.SFiso} and corresponding domain length from \texttt{amep.utils.domain\_length}:} \textbf{a} Isotropic structure factor obtained from the three simulation snapshots shown in Fig.\ \ref{fig:mips} using \texttt{amep.evaluate.SFiso}. The data is smoothed using a running average over seven points with the \texttt{amep.utils.runningmean} function. The inset shows the same data but with logarithmic axes. \textbf{b} Length scale $L(t)$ over time derived from the first moment of the structure factor as defined in Eq.\ (\ref{eq:Lt}) using the \texttt{amep.utils.domain\_length} function. The data has been averaged over five independent ensembles and the black dashed line is a power-law fit of the form $L(t)=L_0t^{\alpha}$ done with \texttt{amep.functions.Fit} resulting in an exponent $\alpha=0.324\pm 0.001$.}
	\label{fig:particle_domain_growth}
\end{figure}

To visually elucidate this process, we present instances of $S(q,t)$ at distinct times $t/\tau_p=0.1, 4.2, 800.0$ (Fig.\ \ref{fig:particle_domain_growth}a) corresponding to the snapshots shown in Fig.\ \ref{fig:mips}. These results can be obtained using the following code example:
\begin{python}
import amep
import numpy as np
# load simulation data
traj = amep.load.traj(
    "/path/to/data",
    mode = "lammps"
)
# calculate S(q)
sf = amep.evaluate.SFiso(
    traj,
    nav = traj.nframes
)
# plot S(q) of the last frame
fig, axs = amep.plot.new()
axs.plot(sf.q, sf.frames[-1][0])
fig.savefig("sf.pdf")
\end{python}
We can now subsequently compute $L(t)$ by using the \pythoninline{amep.utils.domain_length} function and fit a power-law of the form $L(t)=L_0t^{\alpha}$ to it using \pythoninline{amep.functions.Fit}:
\begin{python}[firstnumber=17]
# calculate domain length
L = np.zeros(len(sf.frames))
for i,frame in enumerate(sf.frames):
    length = amep.utils.domain_length(
        frame[0], sf.q, qmax = 0.3
    )
    L[i] = length

# fit growth exponent
def f(x, L0 = 1, alpha = 1):
    return L0*x**alpha
fit = amep.functions.Fit(f)
mask = sf.times > 10
fit.fit(sf.times[mask], L[mask])

# plot data and fit
fig, axs = amep.plot.new()
axs.plot(sf.times, L, label = "data")
axs.plot(
    sf.times[mask],
    fit.generate(sf.times[mask]),
    marker = "", ls = "--",
    label = "fit"
)
axs.legend()
fig.savefig("domain-length.pdf")
\end{python}
The result is illustrated in Fig.\ \ref{fig:particle_domain_growth}b and we obtain a growth exponent of $\alpha=0.324\pm 0.001$ in accordance with the literature \cite{Stenhammar_SoftMatter_2014,Mandal_PhysRevLett_2019,Caporusso_PhysRevLett_2020}.

\subsection{Cluster analysis}
\label{sub:particles-cluster}
Finally, we analyze the clustering of ABPs when undergoing MIPS by using \amep's \pythoninline{cluster} module. For particle-based simulation data, we define a cluster as a collection of particles in which each particle has an interparticle distance smaller than $r_{\rm max}$ to some other particle, where the cut-off distance $r_{\rm max}$ can be chosen by the user. Particle $i$ and $j$ with position coordinates $\vec{r}_i$ and $\vec{r}_j$, respectively, belong to the same cluster if $|\vec{r}_i-\vec{r}_j | \leq r_{\rm max}$. To find the distance between different particle pairs and identify the neighboring particle pairs that satisfy this distance criterion, we use the KDTree algorithm (\pythoninline{scipy.spatial.KDTree}) from the SciPy library \cite{Maneewongvatana_ArXiv_1999,Virtanen_NatMeth_2020}. In the following, we will first show how to identify clusters using \amep\ and afterwards analyze the coarsening of the clusters as well as their radius of gyration and fractal dimension. 

\begin{figure}
	\centering
	\includegraphics[width=1.0\linewidth]{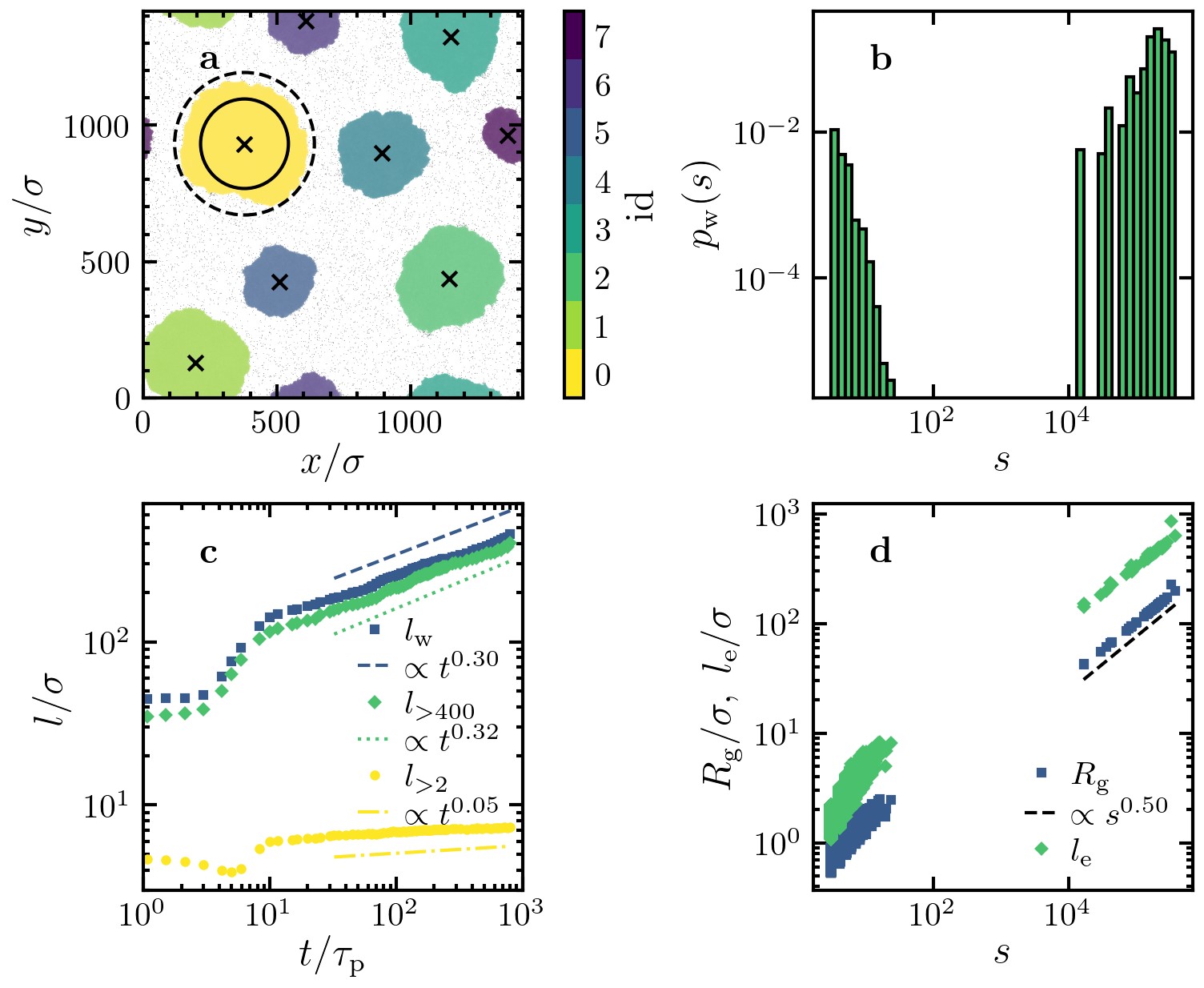}
	\caption{\textbf{Cluster analysis for particle-based simulation data with the \texttt{amep.cluster} and the \texttt{amep.evaluate} modules:} \textbf{a} Same snapshot as in Fig.\ \ref{fig:mips}c showing the eight largest clusters identified with the \texttt{amep.cluster.identify} function and colored with respect to their cluster index. The centers of mass of the clusters calculated with \texttt{amep.cluster.center\_of\_mass} are marked with black crosses. The black solid and dashed circles have radii equal to the radius of gyration and half the linear extension of the largest cluster (yellow) as obtained from Eqs.\ (\ref{eq:Rg}) and (\ref{eq:lin-ext}) using \texttt{amep.cluster.radius\_of\_gyration} and \texttt{amep.cluster.linear\_extension}, respectively. \textbf{b} Weighted cluster size distribution as defined in Eq.\ (\ref{eq:weighted-cdist}) corresponding to the snapshot in panel a as obtained from \texttt{amep.evaluate.ClusterSizeDist} averaged over five independent ensembles. \textbf{c} Cluster length $l/\sigma=\sqrt{\langle s\rangle}$ calculated from the weighted mean cluster size $\langle s\rangle_{\rm w}$ [Eq.\ (\ref{eq:weighted-mean-s})] and mean cluster sizes $\langle s\rangle_{>400}$ and $\langle s\rangle_{>2}$ [Eq.\ \ref{eq:mean-s}] averaged over all clusters larger than 400 and 2 particles, respectively, calculated using \texttt{amep.evaluate.ClusterGrowth} as function of time. The data has been averaged over five independent ensembles. The dashed, dotted, and dash-dotted lines are fits to $l(t)=l(0)t^{\rm \beta}$ done with \texttt{amep.functions.Fit}. \textbf{d} Radius of gyration $R_{\rm g}$ and linear extension $l_{\rm e}$ as function of the cluster size $s$ for the simulation shown in panel a. A fit to $R_{\rm g}(s)\propto s^{1/d_{\rm f}}$ results in a fractal dimension of $d_{\rm f}=1.99\pm 0.03$.}
	\label{fig:particle_cluster}
\end{figure}

Let us consider the last frame of a simulation showing MIPS (Fig.\ \ref{fig:mips}c). The clusters can be identified using the \pythoninline{amep.cluster.identify} function: 
\begin{python}
import amep
import numpy as np
# load simulation data
traj = amep.load.traj(
    "/path/to/data",
    mode = "lammps"
)
# get the last frame
frame = traj[-1]

# cluster detection
clusters, idx = amep.cluster.identify(
    frame.coords(),
    frame.box,
    pbc = True,
    rmax = 1.122
)
\end{python}
It takes the coordinates of all particles (\pythoninline{frame.coords()}) and the boundaries of the simulation box (\pythoninline{frame.box}) as an input and returns a list (\pythoninline{clusters}) of all clusters, sorted in descending order of their size, with each list element containing the indices of the particles belonging to the respective cluster. It also returns the array \pythoninline{idx} which stores the cluster indices assigned to each particle. The optional argument \pythoninline{pbc} is used to consider periodic boundary conditions and the cut-off distance \pythoninline{rmax} is set to the cut-off distance of the WCA potential (i.e., the contact distance). The algorithm works not only with simulation data of particles of the same size but can also identify clusters comprising particles of different sizes, which requires different values of the cut-off distance $r_{\rm max}$ depending on the pair of particles. In the latter case, the user can provide the sizes of the particles (diameter) through the additional keyword \pythoninline{sizes} in \pythoninline{amep.cluster.identify}. In Fig.\ \ref{fig:particle_cluster}a, we show the same snapshot as in Fig.\ \ref{fig:mips}c, but with the particles of the eight largest clusters colored according to their cluster ids.

Based on the identified clusters, one can calculate various properties of the different clusters using the \pythoninline{cluster} module. One of them is the cluster size, defined as the number of particles within a cluster, the distribution of which often provides an insight into coexisting phases in a system \cite{Buttinoni_PhysRevLett_2013,Levis_PhysRevE_2014,Cugliandolo_PhysRevLett_2017,Peruani_PhysRevLett_2012,Peruani_NewJPhys_2013,Liao_SoftMatter_2018,Maloney_SoftMatter_2020,Palaia_JCP_2022}. We can calculate the cluster sizes and masses by using the \pythoninline{amep.cluster.sizes} and \pythoninline{amep.cluster.masses} functions:
\begin{python}[firstnumber=18]
sizes = amep.cluster.sizes(
    clusters
)
masses = amep.cluster.masses(
    clusters, frame.data("mass")
)
\end{python}
Based on \pythoninline{sizes}, we can now calculate the cluster size distribution $p(s)=N_s/\left(\sum_s N_s\right)$, where $N_s$ is the number of clusters with size $s$. Since we are interested in the coarsening behavior of the large clusters but their number is relatively small, it has been proven to be beneficial to analyze the weighted cluster size distribution
\begin{equation}
	p_{\rm w}(s)=\frac{sN_s}{\sum_s sN_s},\label{eq:weighted-cdist}    
\end{equation}
where each cluster is weighted with its size $s$ \cite{Peruani_EPJST_2010,Peruani_PhysRevE_2006}. Here, we use the \pythoninline{amep.statistics.distribution} function with logarithmic bins:
\begin{python}[firstnumber=24]
# weighted cluster size distribution
hist, bins = amep.statistics.distribution(
    sizes, logbins = True,
    nbins = 50, density = False
)
hist = hist*bins/np.sum(hist*bins)

# plot result
fig, axs = amep.plot.new()
axs.bar(
    bins, hist, width = 0.2*bins
)
axs.loglog()
fig.savefig("size-dist.pdf")
\end{python}
The result is shown in Fig.\ \ref{fig:particle_cluster}b and exhibits two distinct peaks at small and large cluster sizes suggesting that the system is phase separated into two distinct phases. Large clusters characterize the dense liquid-like phase whereas the dilute gas-like phase comprises smaller clusters \cite{Peruani_EPJST_2010,Buttinoni_PhysRevLett_2013,Peruani_PhysRevE_2006}.

The cluster sizes can also be used to study the average growth rate of the clusters, which is a commonly used measure of the coarsening kinetics of phases in soft matter and biophysics \cite{Redner_PhysRevLett_2013,Stenhammar_SoftMatter_2014,Caporusso_PhysRevLett_2020,Caporusso_PhysRevLett_2023}. Here, we use \pythoninline{amep.evaluate.ClusterGrowth} to calculate the weighted mean cluster size $\langle s\rangle_{\rm w}$ and the mean cluster size $\langle s\rangle$ defined as
\begin{align}
	\langle s\rangle_{\rm w}&=\sum_s sp_{\rm w}(s)=\frac{\sum_s s^2N_s}{\sum_s sN_s}\label{eq:weighted-mean-s}\\
	\langle s\rangle&=\sum_s sp(s)=\frac{\sum_s sN_s}{\sum_s N_s}\label{eq:mean-s}
\end{align}
over time (Fig.\ \ref{fig:particle_cluster}c). Similar to the domain length $L(t)$ obtained from the structure factor above, we will deduce a growth exponent from the growing mean cluster sizes. Note that for $L(t)$, we only considered $q\sigma <q_{\rm max}\sigma=0.3$, i.e., length scales larger than $l_{\rm min}=2\pi/q_{\rm max}\approx 21\sigma$, which resulted in a growth exponent of $\alpha \approx 1/3$ (where $L(t)\propto t^\alpha$). Since the cluster size measures the number of particles within a cluster, which is proportional to the area of the cluster, we can define a cluster length as $l/\sigma=\sqrt{\langle s\rangle}$ that is expected to also scale as $l\propto t^{1/3}$. However, this holds true only if we either disregard very small clusters that do not grow over time or assign greater importance to larger clusters by calculating the weighted mean cluster size. Therefore, additionally to the weighted mean cluster size, we also compute the mean cluster size considering all the clusters (i.e., having at least two particles) and only the large clusters with at least 400 particles (in accordance to $l_{\rm min}\approx 21\sigma$ used for the calculation of $L(t)$).
\begin{python}[firstnumber=38]
# mean cluster size (>2)
mean_2 = amep.evaluate.ClusterGrowth(
    traj,
    mode = "mean",
    min_size = 2
)
# mean cluster size (>400)
mean_400 = amep.evaluate.ClusterGrowth(
    traj,
    mode = "mean",
    min_size = 400
)
# weighted mean cluster size
weighted_mean = amep.evaluate.ClusterGrowth(
    traj,
    mode = "weighted mean"
)
\end{python}
We now use \pythoninline{amep.functions.Fit} to obtain the growth exponent from $l(t)\propto t^{\beta}$ (here, exemplarily done for the weighted mean in the logarithmic domain):
\begin{python}[firstnumber=55]
# define fit function
def f(l, beta = 1.0, l0 = 1.0):
    return np.log(l0) + beta*l

# create fit object
fit = amep.functions.Fit(f)

# fit function to data at large times
# in logarithmic domain
mask = weighted_mean.times > 3e1
fit.fit(
    np.log(weighted_mean.times[mask]),
    np.log(weighted_mean.frames[mask])/2
)
print(fit.results)

# plot data and fit
fig, axs = amep.plot.new()
axs.plot(
    weighted_mean.times,
    np.sqrt(weighted_mean.frames)
)
axs.plot(
    weighted_mean.times[mask],
    np.exp(fit.generate(
        np.log(weighted_mean.times[mask])
    ))
)
axs.loglog()
fig.savefig("cluster-growth.pdf")
\end{python}
\begin{pythonout}
{'beta': (0.297, 0.004),
 'l0': (64.6, 1.1)}
\end{pythonout}
The results are demonstrated in Fig.\ \ref{fig:particle_cluster}c. Consistent with the established $\propto t^{1/3}$ growth of the characteristic domain length $L(t)$ of such ABP clusters \cite{Stenhammar_SoftMatter_2014}, we obtain an exponent $\beta\approx 1/3$ for the cluster length calculated from the weighted mean and the mean over all clusters that are larger than 400 particles (Fig.\ \ref{fig:particle_cluster}c). For the mean over all clusters larger than two particles, the growth rate is significantly smaller because very small clusters do not grow.

To demonstrate the versatility of the cluster module, we present two additional calculable quantities using \amep: the radius of gyration $R_{\rm g}$ and the linear extension $l_{\rm e}$ of the clusters. For a cluster composed of $s$ particles of masses $m_i, i=1,2, \ldots, s$, located at fixed distances $d_i$ from the cluster's center of mass, the radius of gyration is defined as \cite{Witten_PhysRevLett_1981,Levis_PhysRevE_2014}
\begin{equation}
	R_{\rm g}=\sqrt{\frac{\sum_{i=1}^s m_i d_i^2}{\sum_{i=1}^s m_i}}.\label{eq:Rg}
\end{equation}
The linear extension (also known as the end-to-end length) of a cluster is defined as the maximal distance between two particles in the cluster, i.e,
\begin{equation}
	l_{\rm e}=\max _{\{i, j\}}\left|\vec{r}_i-\vec{r}_j\right|,\label{eq:lin-ext}    
\end{equation}
with $\vec{r}_i$ denoting the position vector of the $i$-th particle and $i,j=1,2, \ldots, s$ \cite{Levis_PhysRevE_2014,Kyriakopoulos_PhysRevE_2019}. The mean linear extension along with the mean size of the largest cluster is commonly used as an order parameter to characterize isotropic percolation phase transitions in different systems \cite{Levis_PhysRevE_2014,Sanoria_PhysRevE_2022,Kyriakopoulos_PhysRevE_2019}. On the other hand, the dependence of the radius of gyration on the cluster size $s$ (or mass) allows us to understand the (possibly fractal \cite{Fehlinger_PhysRevRes_2023}) geometry of the clusters and calculate their fractal dimension $d_{\rm f}$ according to the relation $R_g \propto s^{1/d_{\rm f}}$ \cite{Ghosh_PhysRevLett_2017,Levis_PhysRevE_2014,Johansen_ProteinScience_2011,Brasil_AerosolSciTech_2000,Tentil_PhysRevE_2021,Tokuyama_PhysLettA_1984,Hentschel_PhysRevLett_1984,Paoluzzi_PhysRevE_2018,Fehlinger_PhysRevRes_2023}. With \amep, we can calculate such cluster properties easily by using its \pythoninline{cluster} module (here exemplarily done for the largest cluster):
\begin{python}[firstnumber=85]
# ids of particles in largest cluster
ids = clusters[0]

# center of mass
com = amep.cluster.center_of_mass(
    frame.coords()[ids],
    frame.box,
    frame.data("mass")[ids],
    pbc = True
)
# geometric center
gmc = amep.cluster.geometric_center(
    frame.coords()[ids],
    frame.box,
    pbc = True
)
# radius of gyration
rg = amep.cluster.radius_of_gyration(
    frame.coords()[ids],
    frame.box,
    frame.data("mass")[ids],
    pbc = True
)
# linear extension
le = amep.cluster.linear_extension(
    frame.coords()[ids],
    frame.box,
    frame.data("mass")[ids],
    pbc = True
)
\end{python}
To handle periodic boundary conditions, \amep\ uses the method proposed in Ref.\ \cite{Bai_JGraphTool_2008}. The radius of gyration and the linear extension of the largest cluster are visualized in Fig.\ \ref{fig:particle_cluster}a. Additionally, we plotted $R_{\rm g}$ and $l_{\rm e}$ as function of the cluster size $s$ in Fig.\ \ref{fig:particle_cluster}d to obtain the fractal dimension using \pythoninline{amep.functions.Fit} to fit the function $R_{\rm g}(s)\propto s^{1/d_{\rm f}}$. We obtain $d_{\rm f} = 1.99\pm 0.03$, i.e, essentially the same as the spatial dimension of the system, which is expected here since the clusters are compact.

% =================================================================================================================
% FIELDS
% =================================================================================================================
\section{Analyzing continuum simulation data with \amep}
\label{sec:Fields}
Let us now discuss some examples on how to analyze continuum simulation data with \amep. To this end, we will use numerical solutions of the active model B+, which is a corresponding continuum model for active Brownian particles. We will first introduce the model and afterwards demonstrate and discuss certain observables and minimal code examples using \amep.

\subsection{Active model B+}
\label{sub:AMBP}
The active model B+ (AMB+) describes a system of active Brownian particles featuring a generic self-propulsion mechanism that can lead to interesting collective phenomena. It models the time evolution of the scalar field $\phi=(2\rho-\rho_{\text{H}}-\rho_{\text{L}})/(\rho_{\text{H}}-\rho_{\text{L}})$, where $\rho_{\text{H}}$ and $\rho_{\text{L}}$ denote the particle number density at the low-density and the high-density critical point, respectively \cite{Shaebani_NatRevPhys_2020,Wittkowski_NatComm_2014}. The active model B+ is given by \cite{Tjhung_PhysRevX_2018}
\begin{align}
	\frac{\partial \phi}{\partial t}=&-\nabla\cdot\left[-M\nabla\left(\frac{\delta\mathcal{F}}{\delta\phi}+\lambda|\nabla\phi|^2\right)\right.\nonumber\\
	&\left.+ \zeta M\left(\nabla^2\phi\right)\nabla\phi + \sqrt{2k_{\rm B}TM}\vec{\Lambda}\right], \label{eq:ambp}\\
	\mathcal{F}[ \phi ] =& \int\text{d}^3r\,\left[ \frac{a}{2}\phi^2+\frac{b}{4}\phi^4+\frac{K}{2}|\nabla\phi|^2 \right]. \label{eq:ambp-free-energy}
\end{align}
Here, the free-energy functional $\mathcal{F}$ is approximated up to the order $\phi^4$ and up to square-gradient terms, and $\vec{\Lambda}(\vec{r},t)$ denotes Gaussian white noise with zero mean and unit variance. The mobility is denoted by $M$, the temperature is given by $k_{\rm B}T$ with Boltzmann constant $k_{\rm B}$, and $a,b,K,\zeta,\lambda$ are free parameters of the model. Here, we use $a=-0.25$, $b=0.25$, $K=1.0$, $\zeta=1.0$, $\lambda=-0.5$, $k_{\rm B}T=0.2$, and $M=1.0$ corresponding to a parameter regime showing phase separation \cite{Tjhung_PhysRevX_2018}. We start from a uniform initial condition with $\phi_0=-0.4$ disturbed with weak fluctuations. Here, we used an in-house finite volume solver written in \texttt{C++} to numerically integrate Eq.\ (\ref{eq:ambp}) using a grid of $256\times 256$ grid points with a grid spacing of $\Delta x=\Delta y = 1.0$ and time step $\Delta t=0.001$. All simulations run for $10^8$ time steps. To load the resulting data with \amep, the data has to be stored in a certain data format as discussed in Subsection \ref{sub:field-data-format} and as exemplified in the exemplary dataset which can be downloaded from \url{https://github.com/amepproject/amep/tree/main/examples}.

\subsection{Motility-induced phase separation}
\label{sub:fields-mips}
\begin{figure*}
	\centering
	\includegraphics[width=0.75\linewidth]{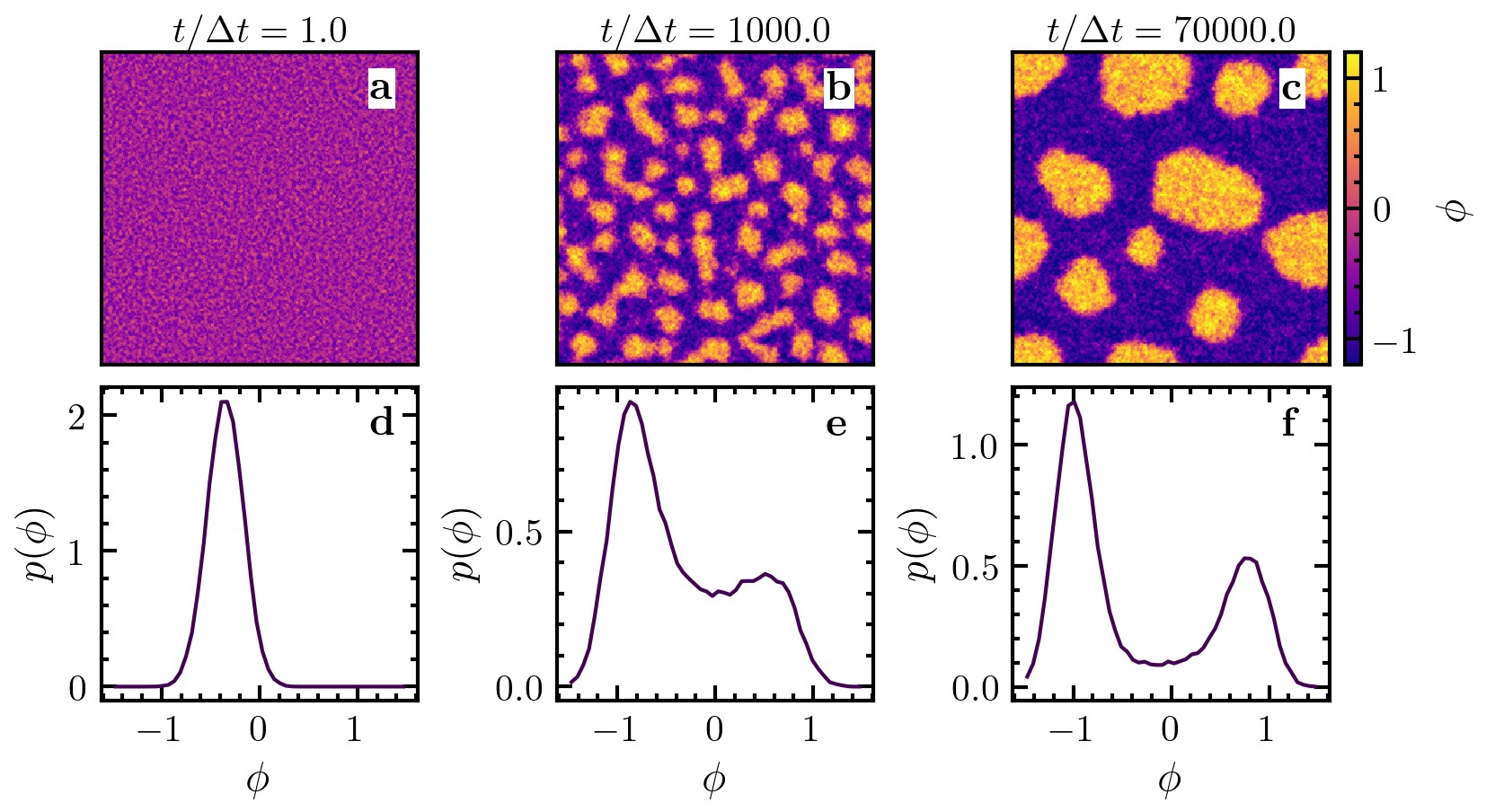}
	\caption{\textbf{Plotting snapshots with \texttt{amep.plot.field} and density distributions using \texttt{amep.evaluate.LDdist}:} \textbf{a--c} Snapshots of numerical solutions of the active model B+ as given in Eq.\ (\ref{eq:ambp}) plotted with \texttt{amep.plot.field} at three different times given in the key. \textbf{d--f} Corresponding density distributions calculated with \texttt{amep.evaluate.LDdist}.}
	\label{fig:fields-snapshots}
\end{figure*}

Similar to the particle-based simulations of ABPs, the AMB+ also undergoes MIPS in a certain parameter regime \cite{Tjhung_PhysRevX_2018}. Whether the system is phase separated can again be determined by evaluating the local density distribution. In case of continuum data, one can simply calculate the distribution of the density at each grid point, as done by the \pythoninline{amep.evaluate.LDdist} class:
\begin{python}
import amep
# load the data
traj = amep.load.traj(
    "/path/to/data",
    mode = "field"
)
# local density distribution of each frame
lddist = amep.evaluate.LDdist(
    traj,
    nav = traj.nframes,
    xmin = -1.5,
    xmax = 1.5,
    ftype = "phi"
)
# save results in HDF5 file
lddist.save("lddist.h5")

# plot for the last frame
fig, axs = amep.plot.new()
axs.plot(lddist.ld, lddist.frames[-1,0])
axs.set_xlabel(r"$\phi_{\rm loc}$")
axs.set_ylabel(r"$p(\phi_{\rm loc})$")
fig.savefig("lddist.pdf")
\end{python}
The results are demonstrated in Fig.\ \ref{fig:fields-snapshots} at different times together with the corresponding snapshots. While the distribution is unimodal at the beginning of the simulations (which start with a uniform distribution), it becomes bimodal at long times showing that the system phase separates into a dense and a dilute phase by forming dense clusters that coexist with a surrounding gas-like phase.

\subsection{Coarsening processes}
\label{sub:fields-sf}
\begin{figure}
	\centering
	\includegraphics[width=1.0\linewidth]{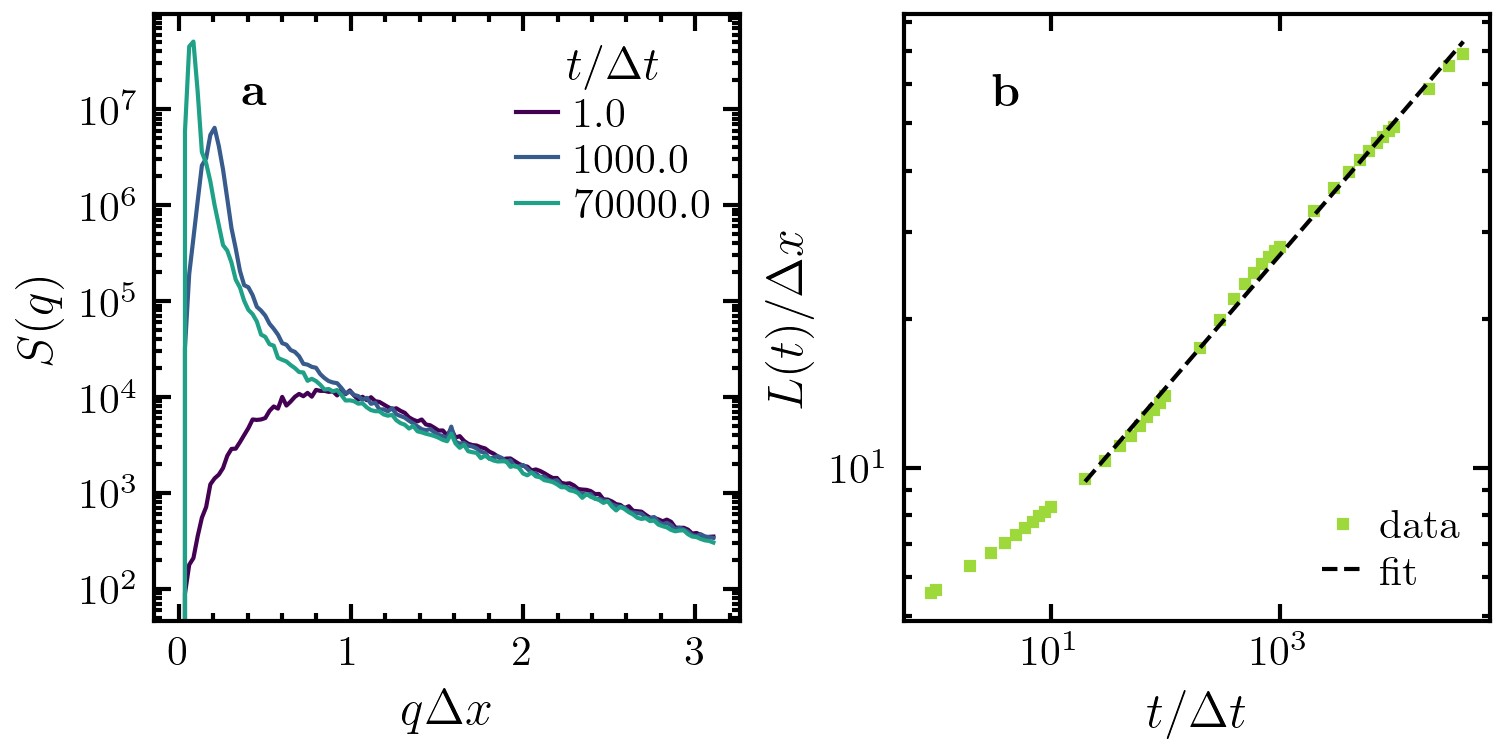}
	\caption{\textbf{Structure factor from \texttt{amep.evaluate.SFiso} and corresponding domain length from \texttt{amep.utils.domain\_length} for continuum simulation data:} \textbf{a} Isotropic structure factor obtained from the three simulation snapshots of the active model B+ shown in Fig.\ \ref{fig:fields-snapshots} calculated with \texttt{amep.evaluate.SFiso}. Note that the upper limit for the wave vector $q$ is given by the resolution of the discretized grid, i.e., $q_{\rm max}=\pi/\Delta x$. \textbf{b} Length scale $L(t)$ as obtained from Eq.\ (\ref{eq:Lt}) using the \texttt{amep.utils.domain\_length} function. The data has been averaged over five independent ensembles and the black dashed line is a power-law fit of the form $L(t)=L_0t^{\alpha}$ done with \texttt{amep.functions.Fit} and resulting in a growth exponent of $\alpha=0.270\pm 0.003$.}
	\label{fig:fields-sf}
\end{figure}

Let us now analyze the coarsening behavior of the clusters based on the isotropic structure factor. For a continuum field, the (two-dimensional) structure factor can directly be calculated from the numerical Fourier transform of the scalar field $\phi$ as \cite{Hansen_Book_TheoryOfSimpleLiquids_2006,Wittkowski_NatComm_2014,Bray_AdvPhys_1994}
\begin{equation}
	S(\vec{q},t) = \left\langle\phi(\vec{q},t)\phi(-\vec{q},t)\right\rangle = \left\langle|\phi(\vec{q},t)|^2\right\rangle
	\label{eq:csf2d}
\end{equation}
using the \pythoninline{amep.continuum.sf2d} function. Here, $\phi(\vec{q},t)$ is the spatial Fourier transform of $\phi(\vec{r},t)$. To obtain the isotropic structure factor, the resulting two-dimensional structure factor must be averaged over the direction of $\vec{q}$, i.e., (using polar coordinates)
\begin{equation}
	S(q,t) = \frac{1}{2\pi}\int_0^{2\pi}{\rm d}\varphi\,S((q\cos(\varphi),q\sin(\varphi)), t)
	\label{eq:sq-from-sf2d}
\end{equation}
which can be done in \amep\ with the \pythoninline{amep.utils.sq_from_sf2d} function:
\begin{python}
import amep
import numpy as np
# load data
traj = amep.load.traj(
    "/path/to/data",
    mode = "field"
)
# get last frame
frame = traj[-1]

# 2d structure factor
sf2d, qx, qy = amep.continuum.sf2d(
    frame.data("phi"),
    *frame.grid
)
# isotropic structure factor
sfiso, q = amep.utils.sq_from_sf2d(
    sf2d, qx, qy
)
\end{python}
\amep's \pythoninline{evaluate} module allows to calculate the isotropic structure factor directly from the \pythoninline{traj} object and also performs a time average:
\begin{python}[firstnumber=20]
# isotropic structure factor
# for all frames (time average)
sf = amep.evaluate.SFiso(
    traj,
    nav = traj.nframes,
    ftype = "phi"
)
\end{python}
Here, we can specify which field of the given trajectory should be used via \pythoninline{ftype}. From the structure factor, we can then again calculate the domain length as defined in Eq.\ (\ref{eq:Lt}) via
\begin{python}[firstnumber=27]
# domain length over time
L = np.zeros(traj.nframes)
for i,f in enumerate(sf.frames):
    L[i] = amep.utils.domain_length(
        f[0], f[1]
    )
\end{python}
Note that we do not specify \pythoninline{qmax} here because we integrate over all $q$ up to its largest possible value $q_{\rm max}=\pi/\Delta x$ given by the grid spacing $\Delta x$. The results are demonstrated in Fig.\ \ref{fig:fields-sf}, which shows three exemplary curves of the structure factor at three different times and the domain length over time. From the domain length, we extract the growth exponent by fitting a power law to the domain length at long times in the logarithmic domain:
\begin{python}[firstnumber=33]
# define fit function
def f(t, L0 = 1.0, alpha = 1.0):
    return np.log(L0) + alpha*t
# create fit object
fit = amep.functions.Fit(f)

# fit at large times
mask = traj.times > 1e1
fit.fit(
    np.log(traj.times[mask]),
    np.log(L[mask])
)
print(fit.results)
\end{python}
\begin{pythonout}
{'L0': (4.168, 0.085),
 'alpha': (0.2702, 0.0029)}
\end{pythonout}
Interestingly, for our continuum simulations of the active model B+, we obtain $\alpha\approx 0.27$, which is smaller than the value $\alpha\approx 1/3$ as obtained in our particle-based simulations. Such subdiffusive scaling has been obtained in previous studies of active field theories as well and is typically expected to be an intermediate scaling regime that leads to a $t^{1/3}$ scaling at longer times \cite{Stenhammar_PhysRevLett_2013,Stenhammar_SoftMatter_2014,Wittkowski_NatComm_2014}.

\subsection{Cluster analysis}
\label{sub:fields-cluster}
Next, we will analyze the clusters forming at late times in more detail by performing a similar cluster analysis as done in Subsection \ref{sub:particles-cluster}. For the continuum data, a cluster is defined as a connected region of similar and higher density than the surroundings. \amep\ provides two algorithms to detect clusters in continuum data (i.e., discretized density fields). The standard algorithm is dividing up pixels according to their values relative to a threshold value, i.e., for a scalar continuum field $\phi$, a zero is assigned to all pixels with $\phi \leq a_{\rm thres}(\phi_{\rm max}-\phi_{\rm min})$ and a one to all pixels with $\phi > a_{\rm thres}(\phi_{\rm max}-\phi_{\rm min})$. Here, $a_{\rm thres}\in[0,1]$ denotes the relative threshold and $\phi_{\rm min}$, $\phi_{\rm max}$ are the minimum and maximum values of the continuum field. On the resulting two-valued image, all connected regions are then labeled as a cluster using \pythoninline{skimage.measure.label} \cite{Virtanen_NatMeth_2020,Wu_MedicalImaging_2005,Fiorio_TheoCompSci_1996}. Within \amep, this algorithm is called \pythoninline{"threshold"}. An alternative way for cluster detection is the watershed algorithm \cite{Romera_Zalitz_BookChapter_Watershed_2017}. It is more stable against slowly varying background fields but needs more fine-tuning of parameters to work. Therefore, the watershed algorithm is not the first choice for detecting clusters over time, i.e., for multiple frames of a trajectory, because the parameters may need to be adjusted for each frame individually. In \amep, the \pythoninline{skimage.segmentation.watershed} function is used for the \pythoninline{"watershed"} cluster detection \cite{Virtanen_NatMeth_2020}.

\begin{figure}
	\centering
	\includegraphics[width=1.0\linewidth]{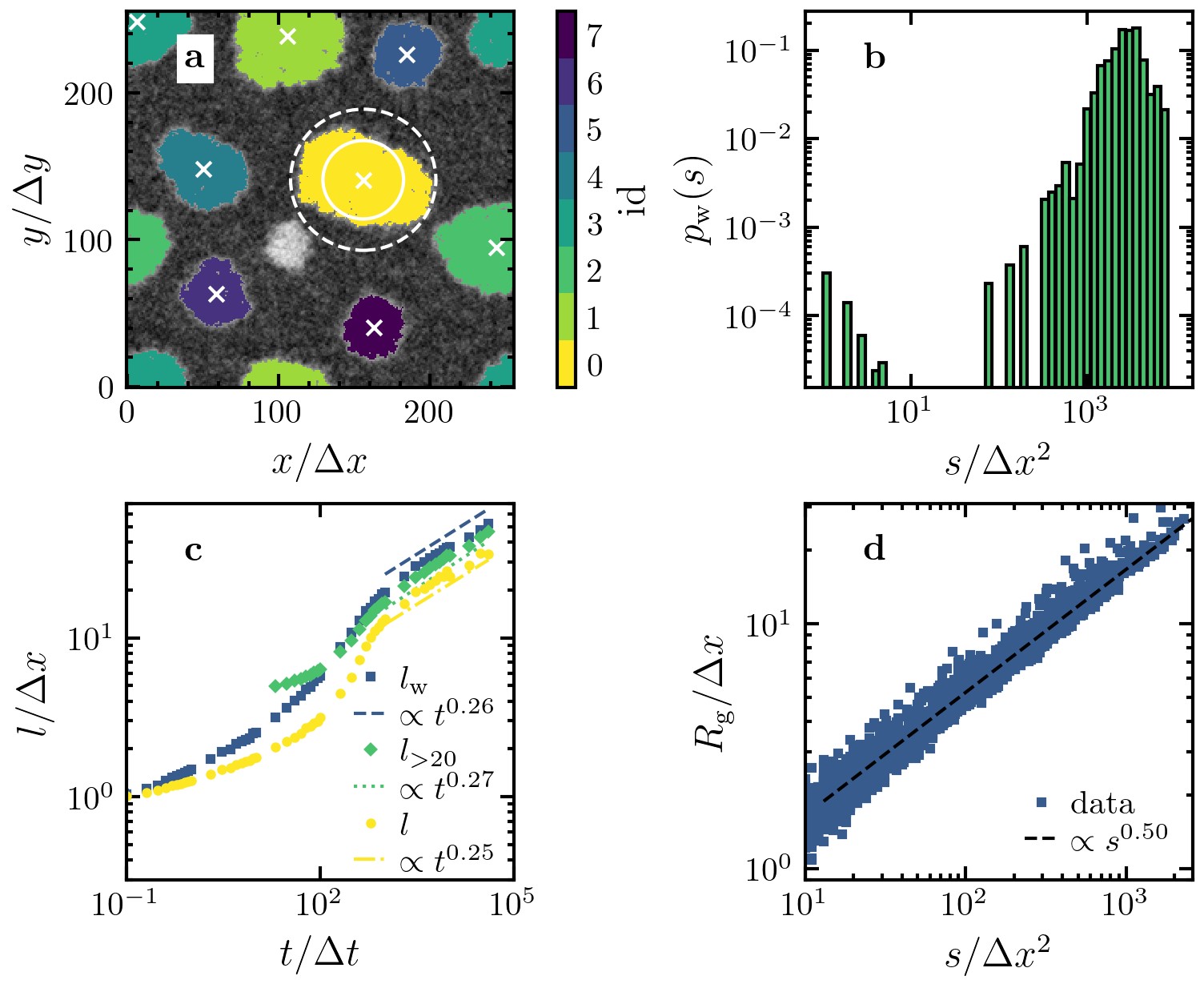}
	\caption{\textbf{Cluster analysis for continuum simulation data with the \texttt{amep.continuum} and the \texttt{amep.evaluate} modules.} \textbf{a} Same snapshot as in Fig.\ \ref{fig:fields-snapshots}c showing the eight largest clusters identified with the \texttt{amep.continuum.identify\_clusters} function and colored with respect to their cluster index. The centers of mass of the clusters calculated with \texttt{amep.continuum.cluster\_properties} are marked with white crosses. The white solid and dashed circles have radii equal to the radius of gyration and half the linear extension of the largest cluster (yellow) as obtained from Eqs.\ (\ref{eq:Rg}) and (\ref{eq:lin-ext}), respectively, using \texttt{amep.continuum.cluster\_properties}. \textbf{b} Weighted cluster size distribution as defined in Eq.\ (\ref{eq:weighted-cdist}) corresponding to the snapshot in panel a as obtained from \texttt{amep.evaluate.ClusterSizeDist} averaged over five independent ensembles. \textbf{c} Cluster length $l=\sqrt{\langle s\rangle}$ obtained from the weighted mean cluster size $\langle s\rangle_{\rm w}$ [Eq.\ (\ref{eq:weighted-mean-s})] and mean cluster sizes $\langle s\rangle$ [Eq.\ \ref{eq:mean-s}] and $\langle s\rangle_{>20}$ averaged over all clusters and over all clusters larger than an area of $20\Delta x^2$, respectively, calculated using \texttt{amep.evaluate.ClusterGrowth} as function of time. The data has been averaged over five independent ensembles. The dashed, dotted, and dash-dotted lines are fits to $l(t)=l(0)t^{\beta}$ done with \texttt{amep.functions.Fit}. \textbf{d} Radius of gyration $R_{\rm g}$ as function of the cluster size $s$ for five independent simulations as exemplarily shown in panel a. A fit to $R_{\rm g}(s)\propto s^{1/d_{\rm f}}$ results in a fractal dimension of $d_{\rm f}=1.99\pm 0.01$.}
	\label{fig:field_cluster_functions}
\end{figure}

Clusters within a continuum field can be identified with \amep\ using the \pythoninline{amep.continuum.identify_clusters} function. In the following example, we detect the clusters in the last frame of a continuum simulation using the \pythoninline{"threshold"} method:
\begin{python}
import amep
# load data
traj = amep.load.traj(
    "path/to/data",
    mode = "field"
)
# get the last frame
frame = traj[-1]

# identify clusters
ids, labels =\
amep.continuum.identify_clusters(
    frame.data("phi"),
    pbc = True,
    threshold = 0.1,
    method = "threshold"
)
\end{python}
Here, \pythoninline{frame.data("phi")} returns an array of values of field \pythoninline{"phi"} at each point of the underlying discretized grid, the \pythoninline{pbc} keyword can be set to \pythoninline{True} to apply periodic boundary conditions, and \pythoninline{threshold} specifies the relative threshold $a_{\rm thres}$. The keyword \pythoninline{method} chooses between the \pythoninline{"threshold"} or \pythoninline{"watershed"} method. The \pythoninline{amep.continuum.identify_clusters} function assigns a unique identifier to each detected cluster returned as NumPy array (\pythoninline{ids} in the example above), and it returns an array of the same shape as the underlying discretized grid denoting which grid point belongs to which cluster (\pythoninline{labels} in the example above). The result is examplarily visualized in Fig.\ \ref{fig:field_cluster_functions}a.

As a next step, the \pythoninline{amep.continuum.cluster_properties} function can be used to calculated certain properties of the detected clusters such as their sizes, masses, centers, or radii of gyration:
\begin{python}[firstnumber=18]
s, gmc, com, rg, le, gt, it =\
amep.continuum.cluster_properties(
    frame.data("phi"),
    *frame.grid,
    ids,
    labels,
    pbc = True
)
\end{python}
where \pythoninline{s, gmc, com, rg, le, gt, it} denote the size, geometric center, center of mass, radius of gyration, linear extension, gyration tensor, and inertia tensor of each cluster, respectively. Here, the size of cluster $i$ within a field $\phi$ is defined as the integral over its area $A_i$, i.e., $s_i=\int_{A_i}{\rm d}^2r\,\phi(\vec{r})$. All other cluster properties are calculated with the same methods as used for the particle-based simulation data by treating each grid point $(i,j)$ as a particle at position $\vec{r}_{ij}$ with ``mass'' $\phi(\vec{r}_{ij})$. The radius of gyration and the linear extension of the largest cluster are visualized in Fig.\ \ref{fig:field_cluster_functions}a.

Similarly to the cluster analysis for particle-based data, we can now calculate the weighted cluster-size distribution, the growth exponents, and the fractal dimension of the clusters exploiting \amep's \pythoninline{evaluate} module. Let us first calculate the weighted cluster-size distribution as defined in Eq.\ (\ref{eq:weighted-cdist}):
\begin{python}[firstnumber=26]
# weighted cluster size distribution
cd = amep.evaluate.ClusterSizeDist(
    traj,
    nav = traj.nframes,
    ftype = "phi",
    method = "threshold",
    threshold = 0.1,
    use_density = False,
    nbins = 50,
    logbins = True,
    xmin = 1e0,
    xmax = 1e4
)
\end{python}
Here, the keyword \pythoninline{use_density} allows to specify whether the integrated density $s_i$ (\pythoninline{use_density = True}) as defined above or area $A_i$ (\pythoninline{use_density = False}) should be used as size of cluster $i$. Furthermore, we have used the threshold method and logarithmic bins and the result is shown in Fig.\ \ref{fig:field_cluster_functions}b. Next, we analyze the cluster growth by calculating the average cluster size over time using the weighted mean and mean as defined in Eqs.\ (\ref{eq:weighted-mean-s}) and (\ref{eq:mean-s}), respectively: 
\begin{python}[firstnumber=39]
# mean cluster sizes
mean = amep.evaluate.ClusterGrowth(
    traj,
    ftype = "phi",
    method = "threshold",
    mode = "mean",
    threshold = 0.1,
    use_density = False
)
# mean cluster size over all clusters
# with size larger than 20
mean_20 = amep.evaluate.ClusterGrowth(
    traj,
    ftype = "phi",
    method = "threshold",
    mode = "mean",
    threshold = 0.1,
    use_density = False,
    min_size = 20
)
# weighted mean cluster size
weighted_mean = amep.evaluate.ClusterGrowth(
    traj,
    ftype = "phi",
    method = "threshold",
    mode = "weighted mean",
    threshold = 0.1,
    use_density = False
)
\end{python}
Again, we define the cluster length as $l=\sqrt{\langle s\rangle}$ and use \pythoninline{amep.functions.Fit} to obtain the growth exponent from $l(t)\propto t^{\beta}$ as already demonstrated in Subsection \ref{sub:particles-cluster}. Consistent with the growth exponent $\alpha\approx 0.27$ obtained from the domain length $L(t)$ (see Subsection \ref{sub:fields-sf}), we obtain $\beta\approx 0.26$ (Fig.\ \ref{fig:field_cluster_functions}c). Additionally, we plotted the radius of gyration $R_{\rm g}$ as function of the cluster size $s$ in Fig.\ \ref{fig:field_cluster_functions}d to obtain the fractal dimension using \pythoninline{amep.functions.Fit} to fit the function $R_{\rm g}(s)\propto s^{1/d_{\rm f}}$. We obtain $d_{\rm f} = 1.99\pm 0.01$, i.e, the same as the spatial dimension of the system, which is expected here since the clusters are compact.

\subsection{Continuum data format}
\label{sub:field-data-format}
Continuum simulation data as analyzed within this work has to be stored in a specific format such that it can be loaded with \amep. In the following, we briefly introduce the basic format requirements. In \amep, field data is internally stored using the \texttt{h5amep} file format, similar to the particle-based simulation data. Converting field data to an \amep\ dataset is done by use of a reader class (\pythoninline{amep.reader.ContinuumReader}). This data reader expects the following format: The standard file structure for field data is inspired by the LAMMPS dump file format, i.e., all relevant data is stored in one base directory which contains (i) a file named \texttt{grid.txt} and (ii) multiple files named \texttt{field\_<index>.txt}. The former is of the form
\begin{verbatim}
BOX:
<X_min> <X_max>
<Y_min> <Y_max>
<Z_min> <Z_max>
SHAPE:
<nx> <ny> <nz>
COORDINATES: X Y Z
<X_0> <Y_0> <Z_0>
<X_1> <Y_0> <Z_0>
...
<X_N> <Y_0> <Z_0>
<X_0> <Y_1> <Z_0>
<X_1> <Y_1> <Z_0>
...
\end{verbatim}
and contains all information about the simulation box and the underlying discrete grid. The values in the \texttt{BOX} category define the borders of the simulation box, which is assumed to be rectangular, the \texttt{SHAPE} category contains the shape of the grid, and \texttt{COORDINATES} contains the coordinates of all grid points. If the data is based on an evenly spaced rectangular grid and the grid points are	given in rising order, the \texttt{SHAPE} category tells \amep\ in what kind of	multidimensional array the data should be cast for representation. The files named \texttt{field\_<index>.txt} contain all data that varies in time. The index should rise in time and could be chosen as the number of timesteps for example. The data files should be of the following form:
\begin{verbatim}
TIMESTEP:
<Simulation timestep>
TIME:
<Physical time>
DATA: <name 0> <name 1> <name 2> <name 3>
<field 0 0> <field 1 0> <field 2 0> <field 3 0>
<field 0 1> <field 1 1> <field 2 1> <field 3 1>
<field 0 2> <field 1 2> <field 2 2> <field 3 2>
...
\end{verbatim}
Here, the names describe the scalar field written to the column beneath, e.g., density. The columns are then filled by the values of the respective field at each grid point in the same order as in \texttt{grid.txt}. The columns can be separated by spaces, tabs, commas, semicolons, colons, or a vertical bar. The \texttt{TIMESTEP} category contains the number of timesteps corresponding to the contained data and the \texttt{TIME} category the corresponding (physical) time. If you have data of this type in the directory \texttt{/path/to/data}, it can be loaded with \amep\ via
\begin{python}
field_trajectory = amep.load.traj(
    "/path/to/data",
    mode = "field",
    delimiter = " ",
    dumps = "field_*.txt",
    reload = True
)
\end{python}
The \pythoninline{mode} keyword tells \amep\ to expect field data, \pythoninline{delimiter} specifies the delimiter used for the columns in the data files, \pythoninline{dumps} takes a regular expression that matches the name format of the variable data files, and the keyword \pythoninline{reload} tells \amep\ whether to look for an existing \texttt{h5amep} file and use it if it exists or to create a new one from the raw data. This should be used when your base data has changed between analysis runs. An exemplary dataset can be downloaded from \url{https://github.com/amepproject/amep/tree/main/examples}.

% =================================================================================================================
% CONCLUSION
% =================================================================================================================
\section{Conclusion}
\label{sec:Conclusion}
\amep\ is a powerful Python library for analyzing simulation data of active matter systems. It provides a unified framework for handling both particle-based and continuum simulation data and combines it with an easy-to-learn Python API. With \amep, one can quickly analyze and plot simulation results and develop new analysis methods utilizing \amep's features and its seamless integrability to powerful scientific Python libraries through NumPy arrays. While its collection of analysis methods is primarily targeted at the active matter community, \amep's general design allows to apply \amep\ to almost any particle-based or continuum simulation data as obtained from classical molecular-dynamics and Brownian-dynamics simulations, or any kind of numerical solutions of partial differential equations.

We have exemplarily shown the potential of \amep\ by analyzing particle-based systems of more than $10^6$ particles and continuum simulations with up to $10^5$ grid points as typically used in active matter research. \amep\ enables the analysis of a broader class of simulation data compared to most other analysis libraries. Such simulation data comprises ``dry'' particle-based models such as the active Brownian particle model and its various relatives, ``wet'' models that explicitly model surrounding fluids including particles that are coupled to flow fields, and continuum models such as the active model B+, the Keller-Segel model, or the Cahn-Hilliard model. Applied to such models, \amep\ can provide essential insights, e.g., into phase separation, pattern formation, and critical phenomena in active matter systems. In addition to the presented observables, many more observables will be included in the future, for instance, entropy production using the method introduced in Ref.\ \cite{Ro_PhysRevLett_2022}, finite-size scaling analysis, and cluster tracking. We would like to encourage the active matter community to contribute to this open-source project in order to complement it with future analysis methods for particle-based and continuum simulation data.

% =================================================================================================================
% ADDITIONAL INFORMATION
% =================================================================================================================
%% author contribution
\section*{CRediT authorship contribution statement}
\textbf{Lukas Hecht:} conceptualization, data curation, formal analysis, funding acquisition, investigation, methodology, project administration, software, validation, visualization, writing (original draft and review \& editing). \textbf{Kay-Robert Dormann:} formal analysis, methodology, software, visualization, writing (original draft and review \& editing). \textbf{Kai Luca Spanheimer:} formal analysis, methodology, software, visualization, writing (original draft and review \& editing). \textbf{Mahdieh Ebrahimi:} formal analysis, software, visualization, writing (original draft and review \& editing). \textbf{Malte Cordts:} software. \textbf{Suvendu Mandal:} formal analysis, investigation, software, visualization, writing (original draft and review \& editing). \textbf{Aritra K.\ Mukhopadhyay:} formal analysis, methodology, software, visualization, writing (original draft and review \& editing). \textbf{Benno Liebchen:} conceptualization, funding acquisition, project administration, supervision, writing (review \& editing).

%% competing interests
\section*{Declaration of competing interest}
The authors declare that they have no known competing financial interests or personal relationships that could have appeared to influence the work reported in this paper.

%% acknowledgements
\section*{Acknowledgements}
L.H.\ and B.L.\ gratefully acknowledge the computing time provided to them at the NHR Center NHR4CES at TU Darmstadt (project number p0020259). This is funded by the Federal Ministry of Education and Research, and the state governments participating on the basis of the resolutions of the GWK for national high performance computing at universities (\url{www.nhr-verein.de/unsere-partner}). L.H.\ gratefully acknowledges the support by the German Academic Scholarship Foundation (Studienstiftung des deutschen Volkes). A.K.M.\ and B.L.\ acknowledge financial support from the Deutsche Forschungsgemeinschaft (DFG, German Research Foundation) through project number 233630050 (TRR-146).

%\bibliography{library_mod}


\end{document}